\documentclass[aps,prb,twocolumn,notitlepage,groupedaddress,showpacs]{revtex4-1}

\newcommand{\altcn}[2]{#2}   

\newcommand{\bgo}{\bar{g}^1}
\newcommand{\bgt}{\bar{g}^2}
\newcommand{\cpo}{{\cal P}^1_{\bf R}}
\newcommand{\cpt}{{\cal P}^2_{\bf R}}
\newcommand{\pfo}{P^1_{\bf R}}
\newcommand{\pft}{P^2_{\bf R}}
\newcommand{\tmo}{t^1_{\bf R}}
\newcommand{\tmt}{t^2_{\bf R}}

\usepackage{graphicx}
\usepackage{bm}
\usepackage{color}

\begin{document}


\title{Nonlocal torque operators in ab initio theory
       of the Gilbert damping in random ferromagnetic alloys}


\author{I. Turek}
\email[]{turek@ipm.cz}
\affiliation{Institute of Physics of Materials,
Academy of Sciences of the Czech Republic,
\v{Z}i\v{z}kova 22, CZ-616 62 Brno, Czech Republic}

\author{J. Kudrnovsk\'y}
\email[]{kudrnov@fzu.cz}
\affiliation{Institute of Physics, 
Academy of Sciences of the Czech Republic,
Na Slovance 2, CZ-182 21 Praha 8, Czech Republic}

\author{V. Drchal}
\email[]{drchal@fzu.cz}
\affiliation{Institute of Physics, 
Academy of Sciences of the Czech Republic,
Na Slovance 2, CZ-182 21 Praha 8, Czech Republic}


\date{\today}

\begin{abstract}
We present an \emph{ab initio} theory of the Gilbert damping
in substitutionally disordered ferromagnetic alloys.
The theory rests on introduced nonlocal torques which
replace traditional local torque operators in the well-known
torque-correlation formula and which can be formulated within
the atomic-sphere approximation.
The formalism is sketched in a simple tight-binding model and
worked out in detail in the relativistic tight-binding linear
muffin-tin orbital (TB-LMTO) method and the coherent potential
approximation (CPA).
The resulting nonlocal torques are represented by nonrandom, 
non-site-diagonal and spin-independent matrices, which simplifies
the configuration averaging.
The CPA-vertex corrections play a crucial role for the internal
consistency of the theory and for its exact equivalence to other
first-principles approaches based on the random local torques.
This equivalence is also illustrated by the calculated Gilbert
damping parameters for binary NiFe and FeCo random alloys,
for pure iron with a model atomic-level disorder,
and for stoichiometric FePt alloys with a varying degree
of L1$_0$ atomic long-range order. 
\end{abstract}


\maketitle


\section{Introduction\label{s_intr}}

The dynamics of magnetization of bulk ferromagnets, utrathin
magnetic films and magnetic nanoparticles represents an important
property of these systems, especially in the context of high speed
magnetic devices for data storage.
While a complete picture of magnetization dynamics including, e.g.,
excitation of magnons and their interaction with other degrees
of freedom, is still a challenge for the modern theory of
magnetism, remarkable progress has been achieved during the last
years concerning the dynamics of the total magnetic moment,
which can be probed experimentally by means of the ferromagnetic
resonance \cite{r_1994_bh} or by the time-resolved magneto-optical
Kerr effect. \cite{r_2014_imn}
Time evolution of the macroscopic magnetization vector ${\bf M}$
can be described by the well-known Landau-Lifshitz-Gilbert (LLG)
equation \cite{r_1980_pl, r_1955_tlg}
\begin{equation}
\frac{d{\bf M}}{dt} =  {\bf B}_{\rm eff} \times {\bf M} 
+ \frac{\bf M}{M} \times 
\left( \underline{\bm{\alpha}} \cdot \frac{d{\bf M}}{dt} \right) ,
\label{eq_llg}
\end{equation}
where ${\bf B}_{\rm eff}$ denotes an effective magnetic field
(with the gyromagnetic ratio absorbed) acting on the magnetization,
$M = |{\bf M}|$, and the quantity $\underline{\bm{\alpha}} = 
\{ \alpha_{\mu\nu} \}$ denotes a symmetric $3 \times 3$ tensor of
the dimensionless Gilbert damping parameters ($\mu, \nu = x, y, z$).
The first term in Eq.~(\ref{eq_llg}) defines a precession of the
magnetization vector around the direction of the effective magnetic
field and the second term describes a damping of the dynamics.
The LLG equation in itinerant ferromagnets is appropriate for
magnetization precessions very slow as compared to precessions of
the single-electron spin due to the exchange splitting and to
frequencies of interatomic electron hoppings.

A large number of theoretical approaches to the Gilbert damping
has been worked out during the last two decades; here we mention
only schemes within the one-electron theory of itinerant magnets,
\cite{r_1976_vk, r_2003_esh, r_2004_sjl, r_2004_tfh, r_2005_sf,
r_2006_fs, r_2007_gis, r_2007_vk, r_2008_btb, r_2008_gis, 
r_2009_gm_a, r_2009_gm_b, r_2010_skb, r_2011_btb, r_2011_emk, 
r_2012_as} where the most important effects of electron-electron
interaction are captured by means of a local spin-dependent
exchange-correlation (XC) potential.
These techniques can be naturally combined with existing
first-principles techniques based on the density-functional theory,
which leads to parameter-free calculations of the Gilbert damping
tensor of pure ferromagnetic metals, their ordered and
disordered alloys, diluted magnetic semiconductors, etc.
One part of these approaches is based on a static limit of the
frequency-dependent spin-spin correlation function of a
ferromagnet. \cite{r_1976_vk, r_2003_esh, r_2004_sjl, r_2004_tfh,
r_2009_gm_a, r_2009_gm_b}
Other routes to the Gilbert damping employ relaxations of
occupation numbers of individual Bloch electron states 
during quasi-static nonequilibrium processes or transition
rates between different states induced by the spin-orbit (SO)
interaction. \cite{r_2005_sf, r_2006_fs, r_2007_gis, r_2007_vk,
r_2008_gis, r_2012_as}
The dissipation of magnetic energy accompanying the slow
magnetization dynamics, evaluated within a scattering theory or the
Kubo linear response formalism, leads also to explicit expressions
for the Gilbert damping tensor. \cite{r_2008_btb, r_2010_skb,
r_2011_btb, r_2011_emk}
Most of these formulations yield relations equivalent to the
so-called torque-correlation formula
\begin{equation}
\alpha_{\mu\nu} = - \alpha_0 {\rm Tr}
\{ T_\mu ( G_+ - G_- ) T_\nu ( G_+ - G_- ) \} ,
\label{eq_tcf}
\end{equation}
in which the torque operators $T_\mu$ are either due to the
XC or SO terms of the one-electron Hamiltonian. 
In Eq.~(\ref{eq_tcf}), which has a form of the Kubo-Greenwood 
formula and is valid for zero temperature of electrons, the 
quantity $\alpha_0$ is related to the system magnetization 
(and to fundamental constants and units used, see 
Section \ref{ss_efft}), the trace is taken over
the whole Hilbert space of valence electrons, and the symbols 
$G_\pm = G(E_{\rm F} \pm {\rm i}0)$ denote the one-particle 
retarded and advanced propagators (Green's functions) at the
Fermi energy $E_{\rm F}$.

Implementation of the above-mentioned theories in first-principles
computational schemes proved opposite trends of the intraband and
interband contributions to the Gilbert damping parameter as functions
of a phenomenological quasiparticle lifetime broadening.
\cite{r_2004_sjl, r_2007_gis, r_2007_vk}
These qualitative studies have recently been put on a more solid
basis by considering a particular mechanism of the lifetime 
broadening, namely, a frozen temperature-induced structural disorder,
which represents a realistic model for a treatment of temperature
dependence of the Gilbert damping. \cite{r_2011_lsy, r_2013_mkw} 
This approach explained quantitatively the low-temperature
conductivity-like and high-temperature resistivity-like trends
of the damping parameters of iron, cobalt and nickel.
Further improvements of the model, including static 
temperature-induced random orientations of local magnetic moments,
have appeared recently. \cite{r_2015_emc}

The \emph{ab initio} studies have also been successful in
reproduction and interpretation of values and concentration
trends of the Gilbert damping in random ferromagnetic
alloys, such as the NiFe alloy with the face-centered cubic (fcc)
structure (Permalloy) \cite{r_2010_skb, r_2013_mkw} and Fe-based
alloys with the body-centered cubic (bcc) structure (FeCo, FeV,
FeSi). \cite{r_2011_emk, r_2011_bmr, r_2013_mkw} 
Other studies addressed also the effects of doping the Permalloy
and bcc iron by $5d$ transition-metal elements \cite{r_2011_emk,
r_2012_as, r_2013_mkw} and of the degree of atomic long-range order
in equiconcentration FeNi and FePt alloys with the L1$_0$-type
structures. \cite{r_2012_as}
Recently, an application to halfmetallic Co-based Heusler alloys
has appeared as well. \cite{r_2015_as_b}
The obtained results revealed correlations of the damping parameter
with the density of states at the Fermi energy and with the 
size of magnetic moments. \cite{r_2011_bmr, r_2013_mkw}

In a one-particle mean-field-like description of a ferromagnet,
the total spin is not conserved due to the XC field and the
SO interaction.
The currently employed forms of the torque operators $T_\mu$ in
the torque-correlation formula (\ref{eq_tcf}) reflect these two
sources; both the XC- and the SO-induced torques are local and
their equivalence for the theory of Gilbert damping has been
discussed by several authors. \cite{r_2009_gm_a, r_2009_gm_b,
r_2015_as_a}
In the case of random alloys, this equivalence rests on a proper
inclusion of vertex corrections in the configuration averaging
of the damping parameters $\alpha_{\mu\nu}$ as two-particle
quantities.

The purpose of the present paper is to introduce another
torque operator that can be used in the torque-correlation
formula (\ref{eq_tcf}) and to discuss its properties. 
This operator is due to intersite electron hopping and it is
consequently nonlocal; in contrast to the local XC- and
SO-induced torques which are random in random crystalline alloys,
the nonlocal torque is nonrandom,
i.e., independent on the particular configuration
of a random alloy, which simplifies the
configuration averaging of Eq.~(\ref{eq_tcf}).
We show that a similar nonlocal effective torque appears in the
fully relativistic linear muffin-tin orbital (LMTO) method in the
atomic-sphere approximation (ASA) used recently for calculations
of the conductivity tensor in spin-polarized random alloys. 
\cite{r_2012_tkd, r_2014_tkd}
Here we discuss theoretical aspects of the averaging in the
coherent-potential approximation (CPA) \cite{r_1969_bv, r_2006_ctk}
and illustrate the developed \emph{ab initio} scheme by applications
to selected binary alloys.
We also compare the obtained results with those of the LMTO-supercell
technique \cite{r_2010_skb} and with other CPA-based techniques, the
fully relativistic Korringa-Kohn-Rostoker (KKR) method 
\cite{r_2011_emk, r_2013_mkw} and the LMTO method with a simplified
treatment of the SO interaction. \cite{r_2012_as}

The paper is organized as follows.
The theoretical formalism is contained in Section \ref{s_form},
with a general discussion of various torque operators and results
of a simple tight-binding model presented in Section \ref{ss_tcf}.
The following Section \ref{ss_efft} describes the derivation of
the LMTO torque-correlation formula with nonlocal torques;
technical details are left to Appendix \ref{app_matrep} concerning
linear-response calculations with varying basis sets and to
Appendix \ref{app_tilt} regarding the LMTO method for systems
with a tilted magnetization direction.
Selected formal properties of the developed theory are discussed
in Section \ref{ss_prop}.
Applications of the developed formalism can be found
in Section \ref{s_illex}.
Details of numerical implementation are listed in 
Section \ref{ss_ind} followed by illustrating examples for systems
of three different kinds: binary solid solutions of 
$3d$ transition metals in Section \ref{ss_bss},
pure iron with a simple model of random potential fluctuations
in Section \ref{ss_fe},
and stoichiometric FePt alloys with a partial long-range order
in Section \ref{ss_fept}.
The main conclusions are summarized in Section \ref{s_conc}.

\section{Theoretical formalism\label{s_form}}

\subsection{Torque-correlation formula with
            alternative torque operators\label{ss_tcf}}

The torque operators $T_\mu$ entering the torque-correlation
formula (\ref{eq_tcf}) are closely related to components of
the time derivative of electron spin.
For spin-polarized systems described by means of an effective 
Schr\"odinger-Pauli one-electron Hamiltonian $H$, acting on
two-component wave functions, the complete time derivative
of the spin operator is given by the commutation relation $t_\mu
= - {\rm i} [ \sigma_\mu / 2 , H ]$, where $\hbar = 1$ is assumed
and $\sigma_\mu$ ($\mu = x , y, z$) denote the Pauli spin matrices.
Let us write the Hamiltonian as $H = H^{\rm p} + H^{\rm xc}$,
where $H^{\rm p}$ includes all spin-independent terms and
the SO interaction (Hamiltonian of a paramagnetic system)
while $H^{\rm xc} = {\bf B}^{\rm xc} ({\bf r}) \cdot \bm{\sigma}$
denotes the XC term due to an effective magnetic field
${\bf B}^{\rm xc} ({\bf r})$.
The complete time derivative (spin torque) can then be written as
$t_\mu = t^{\rm so}_\mu + t^{\rm xc}_\mu$, where
\begin{equation}
t^{\rm so}_\mu = - {\rm i} [ \sigma_\mu / 2 , H^{\rm p} ] ,
\qquad 
t^{\rm xc}_\mu = - {\rm i} [ \sigma_\mu / 2 , H^{\rm xc} ] . 
\label{eq_tsoxc}
\end{equation}
As discussed, e.g., in Ref.~\onlinecite{r_2009_gm_a}, the use of
the complete torque $t_\mu$ in the torque-correlation
formula (\ref{eq_tcf}) leads identically to zero; the correct
Gilbert damping coefficients $\alpha_{\mu\nu}$ follow from
Eq.~(\ref{eq_tcf}) by using either the SO-induced torque 
$t^{\rm so}_\mu$, or the XC-induced torque $t^{\rm xc}_\mu$.
Note that only transverse components (with respect to the easy
axis of the ferromagnet) of the vectors ${\bf t}^{\rm so}$ and
${\bf t}^{\rm xc}$ are needed for the relevant part of the
Gilbert damping tensor (\ref{eq_tcf}). 

The equivalence of both torque operators (\ref{eq_tsoxc}) for the
Gilbert damping can be extended.
Let us consider a simple system described by a model tight-binding
Hamiltonian
$H$, written now as $H = H^{\rm loc} + H^{\rm nl}$, where the first
term $H^{\rm loc}$ is a lattice sum of local atomic-like terms and
the nonlocal second term $H^{\rm nl}$ includes all intersite hopping
matrix elements.
Let us assume that all effects of the SO interaction and XC fields
are contained in the local term $H^{\rm loc}$, so that the hopping
elements are spin-independent and $[ \sigma_\mu , H^{\rm nl} ] = 0$. 
(Note that this assumption, often used in model
studies, is satisfied only approximatively in real ferromagnets with
different widths of the majority and minority spin bands.)
Let us write explicitly $H^{\rm loc} = 
\sum_{\bf R} ( H^{\rm p}_{\bf R} + H^{\rm xc}_{\bf R} )$, where 
${\bf R}$ labels the lattice sites and where $H^{\rm p}_{\bf R}$
comprises the spin-independent part and the SO interaction of the
${\bf R}$th atomic potential while $H^{\rm xc}_{\bf R}$ is due to
the local XC field of the ${\bf R}$th atom.
The operators $H^{\rm p}_{\bf R}$ and $H^{\rm xc}_{\bf R}$ act only
in the subspace of the ${\bf R}$th site; the subspaces of different
sites are orthogonal to each other.
The total spin operator can be written as $\sigma_\mu /2 = (1/2)
\sum_{\bf R} \sigma_{{\bf R} \mu}$, where the local operator
$\sigma_{{\bf R} \mu}$ is the projection of $\sigma_\mu$ on the
${\bf R}$th subspace.
Let us assume that each term $H^{\rm p}_{\bf R}$ is spherically
symmetric and that $H^{\rm xc}_{\bf R} = {\bf B}^{\rm xc}_{\bf R}
\cdot \bm{\sigma}_{\bf R}$, where the effective field
${\bf B}^{\rm xc}_{\bf R}$ of the ${\bf R}$th atom has a constant
size and direction.
Let us introduce local orbital-momentum operators $L_{{\bf R} \mu}$
and their counterparts including the spin, $J_{{\bf R} \mu} =
L_{{\bf R} \mu} + (\sigma_{{\bf R} \mu}/2)$, 
which are generators of local infinitesimal rotations with respect
to the ${\bf R}$th lattice site, and let us define the corresponding
lattice sums $L_\mu = \sum_{\bf R} L_{{\bf R} \mu}$ and 
$J_\mu = \sum_{\bf R} J_{{\bf R} \mu} = L_\mu + (\sigma_\mu/2)$.  
Then the local terms $H^{\rm p}_{\bf R}$ and $H^{\rm xc}_{\bf R}$
satisfy, respectively, commutation rules 
$[ J_{{\bf R} \mu} , H^{\rm p}_{\bf R} ] = 0$ and 
$[ L_{{\bf R} \mu} , H^{\rm xc}_{\bf R} ] = 0$.
By using the above assumptions and definitions, the XC-induced spin
torque (\ref{eq_tsoxc}) due to the XC term $H^{\rm xc} = \sum_{\bf R}
H^{\rm xc}_{\bf R}$ can be reformulated as
\begin{eqnarray}
\label{eq_loct}
t^{\rm xc}_\mu & = & - {\rm i} \sum_{\bf R} 
[ \sigma_{{\bf R} \mu} / 2 , H^{\rm xc}_{\bf R} ] 
 = - {\rm i} \sum_{\bf R} [ J_{{\bf R} \mu} , H^{\rm xc}_{\bf R} ]
\altcn{ \nonumber\\ }{ \\ }
 & = & - {\rm i} \sum_{\bf R} 
[ J_{{\bf R} \mu} , H^{\rm p}_{\bf R} + H^{\rm xc}_{\bf R} ]
 = - {\rm i} [ J_\mu , H^{\rm loc} ] 
\equiv t^{\rm loc}_\mu .
\altcn{ }{ \nonumber }
\end{eqnarray}
The last commutator defines a local torque operator $t^{\rm loc}_\mu$
due to the local part of the Hamiltonian $H^{\rm loc}$ and the 
operator $J_\mu$, in contrast to the spin operator $\sigma_\mu/2$ in
Eq.~(\ref{eq_tsoxc}).
Let us define the complementary nonlocal torque $t^{\rm nl}_\mu$ due
to the nonlocal part of the Hamiltonian $H^{\rm nl}$, namely,
\begin{equation}
t^{\rm nl}_\mu = - {\rm i} [ J_\mu , H^{\rm nl} ] 
= - {\rm i} [ L_\mu , H^{\rm nl} ] ,
\label{eq_nlt}
\end{equation}
and let us employ the fact that the complete time derivative of
the operator $J_\mu$, i.e., the torque $\tilde{t}_\mu = 
- {\rm i} [ J_\mu , H ] = t^{\rm loc}_\mu + t^{\rm nl}_\mu$, leads
identically to zero when used in Eq.~(\ref{eq_tcf}).
This fact implies that the Gilbert damping parameters can be also 
obtained from the torque-correlation formula with the nonlocal
torques $t^{\rm nl}_\mu$.
These torques are equivalent to the original
spin-dependent local XC- or SO-induced torques; however,
the derived nonlocal torques are spin-independent, so that
commutation rules $[ t^{\rm nl}_\mu , \sigma_\nu ] = 0$ are
satisfied.

In order to see the effect of different forms of the torque
operators, Eqs.~(\ref{eq_tsoxc}) and (\ref{eq_nlt}), we have
studied a tight-binding model of $p$-orbitals on a simple cubic
lattice with the ground-state magnetization along $z$ axis.
The local (atomic-like) terms of the Hamiltonian are specified
by the XC term $b\sigma_{{\bf R}z}$ and the SO term 
$\xi {\bf L}_{\bf R} \cdot \bm{\sigma}_{\bf R}$, which are added
to a random spin-independent $p$-level at energy $\epsilon_0 + 
D_{\bf R}$, where $\epsilon_0$ denotes the nonrandom center of
the $p$-band while the random parts $D_{\bf R}$ satisfy
configuration averages $\langle D_{\bf R} \rangle = 0$ and
$\langle D_{{\bf R}'} D_{\bf R} \rangle = \gamma 
\delta_{{\bf R}' {\bf R}}$ with the disorder strength $\gamma$.
The spin-independent nonlocal (hopping) part of the Hamiltonian
has been confined to nonrandom nearest-neighbor hoppings
parametrized by two quantities, $W_1$ ($pp\sigma$ hopping) and
$W'_1$ ($pp\pi$ hopping), see, e.g., page 36 of 
Ref.~\onlinecite{r_1992_ag}.
The particular values have been set to $b=0.3$, $\xi=0.2$,
$E_{\rm F} - \epsilon_0 = 0.1$, $\gamma=0.05$, $W_1 = 0.3$ and
$W'_1=-0.1$ (the hoppings were chosen such that the band edges
for $\epsilon_0=b=\xi=\gamma=0$ are $\pm 1$).
The configuration average of the propagators
$\langle G_\pm \rangle = \bar{G}_\pm$ and of the torque
correlation (\ref{eq_tcf}) was performed in the self-consistent
Born approximation (SCBA) including the vertex corrections. 
Since all three torques, Eqs.~(\ref{eq_tsoxc}) and (\ref{eq_nlt}),
are nonrandom operators in our model, the only relevant component
of the Gilbert damping tensor, namely $\alpha_{xx} = \alpha_{yy} 
= \alpha$, could be unambiguously decomposed in the coherent part 
$\alpha^{\rm coh}$ and the incoherent part 
$\alpha^{\rm vc}$ due to the vertex corrections. 

\begin{figure}
\altcn{ \includegraphics[width=0.55\textwidth]{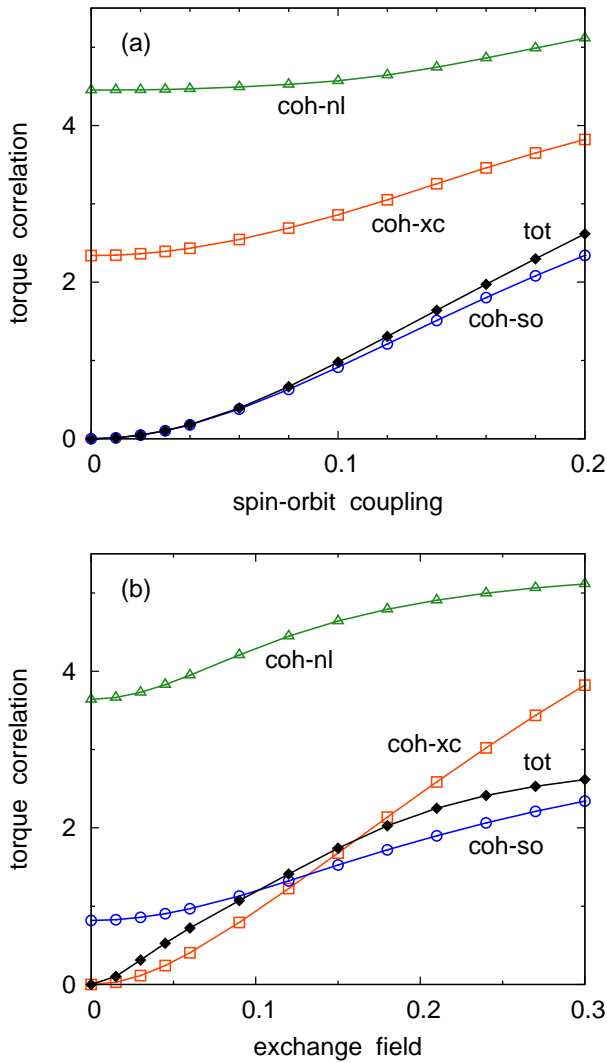} }%
{ \includegraphics[width=0.95\columnwidth]{fig_s1} }
\caption{(Color online)
The torque correlation $\alpha/\alpha_0$, Eq.~(\ref{eq_tcf}),
in a tight-binding $p$-orbital model treated in the SCBA
as a function of the spin-orbit coupling $\xi$ (a) and
of the exchange field $b$ (b).
The full diamonds display the total torque correlation (tot)
and the open symbols denote the coherent contributions 
$\alpha^{\rm coh}/\alpha_0$ calculated with
the SO-induced torque (coh-so), the XC-induced torque (coh-xc), 
Eq.~(\ref{eq_tsoxc}), and the nonlocal torque (coh-nl), 
Eq.~(\ref{eq_nlt}).
\label{f_scba}}
\end{figure}

The results are summarized in Fig.~\ref{f_scba} which displays the
torque correlation $\alpha/\alpha_0$ as a function of the
SO coupling $\xi$ (Fig.~\ref{f_scba}a) and the XC field $b$ 
(Fig.~\ref{f_scba}b).
The total value $\alpha = \alpha^{\rm coh} + 
\alpha^{\rm vc}$ is identical for all three forms of the
torque operator, in contrast to the coherent parts 
$\alpha^{\rm coh}$ which exhibit markedly different values
and trends as compared to each other and to the total $\alpha$.
This result is in line with conclusions drawn by the authors of 
Ref.~\onlinecite{r_2009_gm_a, r_2009_gm_b, r_2015_as_a} proving
the importance of the vertex corrections for obtaining the same
Gilbert damping parameters from the SO- and XC-induced torques.
The only exception seems to be the case of the SO splitting much
weaker than the exchange splitting, where the vertex corrections
for the SO-induced torque can be safely neglected, see 
Fig.~\ref{f_scba}a.
This situation, encountered in $3d$ transition metals and their
alloys, has been treated with the SO-induced torque on an
\emph{ab initio} level with neglected vertex corrections in
Ref.~\onlinecite{r_2007_gis, r_2007_vk}. 
On the other hand, the use of the XC-induced torque calls for
a proper evaluation of the vertex corrections; their neglect
leads to quantitatively and physically incorrect results as
documented by recent first-principles studies. 
\cite{r_2011_emk, r_2013_mkw}
The vertex corrections are indispensable also for the nonlocal
torque, in particular for correct vanishing of the total torque
correlation both in the nonrelativistic limit ($\xi \to 0$, 
Fig.~\ref{f_scba}a) and in the nonmagnetic limit ($b \to 0$, 
Fig.~\ref{f_scba}b).

Finally, let us discuss briefly the general equivalence of the
SO- and XC-induced spin torques, Eq.~(\ref{eq_tsoxc}), in the
fully relativistic four-component Dirac formalism. \cite{r_2005_zhs,
r_2011_ekm}
The Kohn-Sham-Dirac Hamiltonian can be written as $H = H^{\rm p}
+ H^{\rm xc}$, where $H^{\rm p} = c \bm{\alpha} \cdot {\bf p}
+ mc^2 \beta + V ({\bf r})$ and $H^{\rm xc} = 
{\bf B}^{\rm xc} ({\bf r}) \cdot \beta \bm{\Sigma}$, where
$c$ is the speed of light, $m$ denotes the electron mass,
${\bf p} = \{ p_\mu \}$ refers to the momentum operator,
$V ({\bf r})$ is the spin-independent part of the effective
potential and the $\bm{\alpha} = \{ \alpha_\mu \}$, $\beta$ and
$\bm{\Sigma} = \{ \Sigma_\mu \}$ are the well-known $4 \times 4$
matrices of the Dirac theory. \cite{r_1961_mer, r_1998_ps}
Then the XC-induced torque is ${\bf t}^{\rm xc} = 
{\bf B}^{\rm xc} ({\bf r}) \times \beta \bm{\Sigma}$, which is
currently used in the KKR theory of the Gilbert damping.
\cite{r_2011_emk, r_2013_mkw}
The SO-induced torque is ${\bf t}^{\rm so} = {\bf p} \times 
c \bm{\alpha}$, i.e., it is given directly by the relativistic
momentum (${\bf p}$) and velocity ($c \bm{\alpha}$) operators. 
One can see that the torque ${\bf t}^{\rm so}$ is local but
independent of the particular system studied.
A comparison of both alternatives, concerning the total damping
parameters as well as their coherent and incoherent parts, would be
desirable; however, this task is beyond the scope of the present
study.

\subsection{Effective torques in the LMTO method\label{ss_efft}}

In our \emph{ab initio} approach to the Gilbert damping, we
employ the torque-correlation formula (\ref{eq_tcf}) with torques
derived from the XC field. \cite{r_2009_gm_a, r_2011_emk,
r_2013_mkw}
The torque operators are constructed by considering infinitesimal
deviations of the direction of the XC field of the ferromagnet from
its equilibrium orientation, taken as a reference state.
These deviations result from rotations by small angles around axes
perpendicular to the equilibrium direction of the XC field; 
components of the torque operator are then given as derivatives
of the one-particle Hamiltonian with respect to the rotation
angles. \cite{r_2009_ct}

For practical evaluation of Eq.~(\ref{eq_tcf}) in an \emph{ab initio}
technique (such as the LMTO method), one has to consider a matrix
representation of all operators in a suitable orthonormal basis.
The most efficient techniques of the electronic structure theory
require typically basis vectors tailored to the system studied;
in the present context, this leads naturally to basis sets depending
on the angular variables needed to define the torque operators.
Evaluation of the torque correlation using angle-dependent bases
is discussed in Appendix \ref{app_matrep}, where we prove that
Eq.~(\ref{eq_tcf}) can be calculated solely from the matrix elements
of the Hamiltonian and their angular derivatives, see 
Eq.~(\ref{eq_app_alfafin}), whereas the angular dependence of the
basis vectors does not contribute directly to the final result.

The relativistic LMTO-ASA Hamiltonian matrix for the reference
system in the orthogonal LMTO representation is given by
\cite{r_1991_slg, r_1996_sdk, r_1997_tdk}
\begin{equation}
H = C + (\sqrt{\Delta})^+ S ( 1 - \gamma S )^{-1} \sqrt{\Delta} ,
\label{eq_hlmto}
\end{equation}
where the $C$, $\sqrt{\Delta}$ and $\gamma$ denote site-diagonal
matrices of the standard LMTO potential parameters and $S$ is
the matrix of canonical structure constants.
The change of the Hamiltonian matrix $H$ due to a uniform rotation
of the XC field is treated in Appendix \ref{app_tilt};
it is summarized for finite rotations in Eq.~(\ref{eq_app_hortf})
and for angular derivatives of $H$ in Eqs.~(\ref{eq_app_dhdt})
and (\ref{eq_app_dhdto}).
The resolvent $G(z) = (z-H)^{-1}$ of the LMTO Hamiltonian
(\ref{eq_hlmto}) for complex energies $z$ can be expressed
using the auxiliary resolvent $g(z) = [ P(z) - S ]^{-1}$,
which represents an LMTO-counterpart of the scattering-path
operator matrix of the KKR method. \cite{r_2005_zhs, r_2011_ekm}
The symbol $P(z)$ denotes the site-diagonal matrix of potential
functions; their analytic dependence on $z$ and on the potential
parameters can be found elsewhere. \cite{r_1991_slg, r_2012_tkd}
The relation between both resolvents leads to the formula
\cite{r_2014_tkd}
\begin{equation}
G_+ - G_- = F ( g_+ - g_- ) F^+ ,
\label{eq_gg}
\end{equation}
where the same abbreviation $F = (\sqrt{\Delta})^{-1} 
(1 - \gamma S)$ as in Eq.~(\ref{eq_app_dhdt}) was used and 
$g_\pm = g(E_{\rm F} \pm {\rm i}0)$ .

The torque-correlation formula (\ref{eq_tcf}) in the LMTO-ASA method
follows directly from relations (\ref{eq_app_alfafin}), 
(\ref{eq_app_dhdt}), (\ref{eq_app_dhdto}) and (\ref{eq_gg}).
The components of the Gilbert damping tensor $\{ \alpha_{\mu\nu} \}$
in the LLG equation (\ref{eq_llg}) can be obtained from a basic
tensor $\{ {\tilde \alpha}_{\mu\nu} \}$ given by
\begin{equation}
{\tilde \alpha}_{\mu\nu} = - \alpha_0 \, {\rm Tr} \{
\tau_\mu ( g_+ - g_- ) \tau_\nu ( g_+ - g_- ) \} ,
\label{eq_talpha}
\end{equation}
where the quantities 
\begin{equation}
\tau_\mu = - {\rm i} [ {\cal J}^\mu , S ] 
         = - {\rm i} [ {\cal L}^\mu , S ] 
\label{eq_efftq}
\end{equation}
define components of an effective torque in the LMTO-ASA method.
The site-diagonal matrices ${\cal J}^\mu$ and ${\cal L}^\mu$
($\mu = x , y , z$) are Cartesian components of the total and
orbital angular momentum operator, respectively, see text around
Eqs.~(\ref{eq_app_dhdt}) and (\ref{eq_app_dhdto}).
The trace in (\ref{eq_talpha}) extends over all orbitals of the
crystalline solid and the prefactor can be written as $\alpha_0
= (2\pi M_{\rm spin})^{-1}$, where $M_{\rm spin}$ denotes
the spin magnetic moment of the whole crystal in units of the
Bohr magneton $\mu_{\rm B}$. \cite{r_2009_gm_a, r_2011_emk, 
r_2013_mkw}

Let us discuss properties of the effective torque (\ref{eq_efftq}).
Its form is obviously identical to the nonlocal torque
(\ref{eq_nlt}).
The matrix $\tau_\mu$ is non-site-diagonal, but---for a random
substitutional alloy on a nonrandom lattice---it is nonrandom
(independent on the alloy configuration).
Moreover, it is given by a commutator of the site-diagonal
nonrandom matrix ${\cal J}^\mu$ (or ${\cal L}^\mu$) and the LMTO
structure-constant matrix $S$.
These properties point to a close analogy between the effective
torque and the effective velocities in the LMTO conductivity tensor
based on a concept of intersite electron hopping.
\cite{r_2002_tkd, r_2012_tkd, r_2014_tkd}
Let us mention that existing \emph{ab initio} approaches employ
random torques, either the XC-induced torque in the KKR method
\cite{r_2011_emk, r_2013_mkw} or the SO-induced torque in the LMTO
method. \cite{r_2012_as}
Another interesting property of the effective torque $\tau_\mu$
(\ref{eq_efftq}) is its spin-independence which follows from the
spin-independence of the matrices ${\cal L}^\mu$ and $S$.

The explicit relation between the symmetric tensors 
$\{ \alpha_{\mu\nu} \}$ and $\{ {\tilde \alpha}_{\mu\nu} \}$
can be easily formulated for the ground-state magnetization along
$z$ axis; then it is given simply by 
$\alpha_{xx} = {\tilde \alpha}_{yy}$, 
$\alpha_{yy} = {\tilde \alpha}_{xx}$, 
and $\alpha_{xy} = - {\tilde \alpha}_{xy}$.
These relations reflect the fact that an infinitesimal deviation
towards $x$ axis results from an infinitesimal rotation of the
magnetization vector around $y$ axis and vice versa.
Note that the other components of the Gilbert damping tensor 
($\alpha_{\mu z}$ for $\mu =  x, y, z$) are not relevant for the
dynamics of small deviations of magnetization direction described
by the LLG equation (\ref{eq_llg}). 
For the ground-state magnetization pointing along a general
unit vector ${\bf m} = ( m_x , m_y , m_z )$, one has to employ the
Levi-Civita symbol $\epsilon_{\mu\nu\lambda}$ in order to get the
Gilbert damping tensor $\underline{\bm{\alpha}}$ as 
\begin{equation}
\alpha_{\mu\nu} = \sum_{\mu' \nu'} 
\eta_{\mu \mu'} \eta_{\nu \nu'} {\tilde \alpha}_{\mu' \nu' } ,
\label{eq_altal}
\end{equation}
where 
$\eta_{\mu \nu} = \sum_\lambda \epsilon_{\mu\nu\lambda} m_\lambda$.
The resulting tensor $(\ref{eq_altal})$ satisfies the condition
$\underline{\bm{\alpha}} \cdot {\bf m} = 0$ appropriate for
the dynamics of small transverse deviations of magnetization.

The application to random alloys requires configuration averaging
of ${\tilde \alpha}_{\mu\nu}$ (\ref{eq_talpha}).
Since the effective torques $\tau_\mu$ are nonrandom, one can
write a unique decomposition of the average into the coherent
and incoherent parts, ${\tilde \alpha}_{\mu\nu} = {\tilde 
\alpha}^{\rm coh}_{\mu\nu} + {\tilde \alpha}^{\rm vc}_{\mu\nu}$,
where the coherent part is expressed by means of the averaged
auxiliary resolvents ${\bar g}_\pm = \langle g_\pm \rangle$ as
\begin{equation}
{\tilde \alpha}^{\rm coh}_{\mu\nu} = - \alpha_0 {\rm Tr} \{
\tau_\mu ( {\bar g}_+ - {\bar g}_- ) 
\tau_\nu ( {\bar g}_+ - {\bar g}_- ) \} 
\label{eq_talpha_coh}
\end{equation}
and the incoherent part (vertex corrections) is given as
a sum of four terms, namely,
\begin{equation}
{\tilde \alpha}^{\rm vc}_{\mu\nu} = - \alpha_0 
\sum_{p = \pm} \sum_{q = \pm} {\rm sgn} (pq) {\rm Tr} \langle
\tau_\mu g_p \tau_\nu g_q \rangle_{\rm vc} .
\label{eq_talpha_vc}
\end{equation}
In this work, the configuration averaging has been done in the CPA.
Details concerning the averaged resolvents can be found, e.g.,
in Ref.~\onlinecite{r_1997_tdk} and the construction of the vertex
corrections for transport properties was described in Appendix to 
Ref.~\onlinecite{r_2006_ctk}.

\subsection{Properties of the LMTO torque-correlation
            formula\label{ss_prop}} 

The damping tensor (\ref{eq_talpha}) has been formulated in the
canonical LMTO representation.
In the numerical implementation, the well-known transformation
to a tight-binding (TB) LMTO representation \cite{r_1984_aj,
r_1986_apj} is advantageous.
The TB-LMTO representation is specified by a diagonal
matrix $\beta$ of spin-independent screening constants
($\beta_{{\bf R}' \ell' m' s', {\bf R} \ell m s} 
= \delta_{{\bf R}' {\bf R}} \delta_{\ell' \ell} \delta_{m' m} 
\delta_{s' s} \beta_{{\bf R}\ell}$ in a nonrelativistic basis)
and the transformation of all quantities between both LMTO
representations has been discussed in the literature for pure
crystals \cite{r_1986_apj} as well as for random alloys.
\cite{r_1997_tdk, r_2000_tkd, r_2014_tkd}
The same techniques can be used in the present case together with
an obvious commutation rule 
$[ {\cal J}^\mu , \beta ] = [ {\cal L}^\mu , \beta ] = 0$.
Consequently, the conclusions drawn are the same as for the 
conductivity tensor: \cite{r_2014_tkd} 
the total damping tensor (\ref{eq_talpha}) as
well as its coherent (\ref{eq_talpha_coh}) and incoherent 
(\ref{eq_talpha_vc}) parts in the CPA are invariant with respect
to the choice of the LMTO representation.

It should be mentioned that the central result, namely the relations
(\ref{eq_talpha}) and (\ref{eq_efftq}), is not limited to the
LMTO theory, but it can be translated into the KKR theory as well,
similarly to the conductivity tensor in the formalism of intersite
hopping. \cite{r_2002_tkd} 
The LMTO structure-constant matrix $S$ and the auxiliary Green's
function $g(z)$ will be then replaced respectively by the KKR
structure-constant matrix and by the scattering-path operator.
\cite{r_2005_zhs, r_2011_ekm}
Note, however, that the total (${\cal J}^\mu$) and orbital
(${\cal L}^\mu$) angular momentum operators in the effective
torques (\ref{eq_efftq}) will be represented by the same matrices
as in the LMTO theory.

Let us mention for completeness that the present LMTO-ASA theory
allows one to introduce effective local (but random) torques as well.
This is based on the fact that only the Fermi-level propagators
$g_\pm$ defined by the structure constant matrix $S$ and by the
potential functions at the Fermi energy, $P = P(E_{\rm F})$, enter
the zero-temperature expression for the damping tensor
${\tilde \alpha}_{\mu\nu}$ (\ref{eq_talpha}).
Since the equation of motion $(P - S) g_\pm = 1$ implies immediately
$S ( g_+ - g_- ) = P ( g_+ - g_- )$ and, similarly,
$( g_+ - g_- ) S = ( g_+ - g_- ) P$, one can obviously replace the
nonlocal torques $\tau_\mu$ (\ref{eq_efftq}) in the
torque-correlation formula (\ref{eq_talpha}) by their local
counterparts 
\begin{equation}
\tau^{\rm xc}_\mu = {\rm i} [ P , {\cal J}^\mu ] , \qquad
\tau^{\rm so}_\mu = {\rm i} [ P , {\cal L}^\mu ] .
\label{eq_effloctq}
\end{equation}
These effective torques are represented by random, site-diagonal
matrices; the $\tau^{\rm xc}_\mu$ and $\tau^{\rm so}_\mu$
correspond, respectively, to the XC-induced torque used in the
KKR method \cite{r_2013_mkw} and to the SO-induced torque used
in the LMTO method with a simplified treatment of the
SO-interaction. \cite{r_2012_as} 
In the case of random alloys treated in the CPA, the randomness
of the local torques (\ref{eq_effloctq}) calls for the approach
developed by Butler \cite{r_1985_whb} for the averaging of the
torque-correlation coefficient (\ref{eq_talpha}).
One can prove that the resulting damping parameters
${\tilde \alpha}_{\mu\nu}$ obtained in the CPA with the local and
nonlocal torques are fully equivalent to each other;
this equivalence rests heavily on a proper inclusion of the
vertex corrections \cite{r_sumagd} and it leads to further
important consequences.
First, the Gilbert damping tensor vanishes exactly for zero 
SO interaction, which follows from the use of the SO-induced
torque $\tau^{\rm so}_\mu$ and from the obvious commutation rule
$[ P, {\cal L}^\mu ] = 0$ valid for the spherically symmetric
potential functions (in the absence of SO interaction).
This result is in agreement with the numerical study of the toy
model in Section~\ref{ss_tcf}, see Fig.~\ref{f_scba}a for $\xi = 0$.
On an \emph{ab initio} level, this property has been obtained
numerically both in the KKR method \cite{r_2013_mkw} and in the LMTO
method. \cite{r_2015_as_a}
Second, the XC- and SO-induced local torques (\ref{eq_effloctq})
within the CPA are exactly equivalent as well,
as has been indicated in a recent numerical study for a random
bcc Fe$_{50}$Co$_{50}$ alloy. \cite{r_2015_as_a}
In summary, the nonlocal torques (\ref{eq_efftq}) and both local
torques (\ref{eq_effloctq}) can be used as equivalent alternatives
in the torque-correlation formula (\ref{eq_talpha}) provided that
the vertex corrections are included consistently with the
CPA-averaging of the single-particle propagators.

\section{Illustrating examples\label{s_illex}}

\subsection{Implementation and numerical details\label{ss_ind}} 

\begin{figure}
\altcn{ \includegraphics[width=0.55\textwidth]{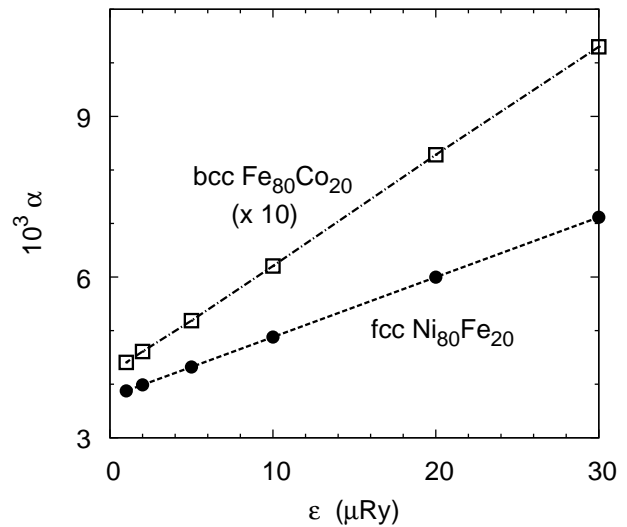} }%
{ \includegraphics[width=0.95\columnwidth]{fig_s2} }
\caption{The Gilbert damping parameters $\alpha$ of random fcc
Ni$_{80}$Fe$_{20}$ (full circles) and bcc Fe$_{80}$Co$_{20}$ 
(open squares) alloys as functions of the imaginary part of
energy $\varepsilon$.
The values of $\alpha$ for the Fe$_{80}$Co$_{20}$ alloy are
magnified by a factor of 10.
\label{f_alpeps}}
\end{figure}

The numerical implementation of the described theory and the
calculations have been done with similar tools as in our recent
studies of ground-state \cite{r_2012_tkc} and transport 
\cite{r_2012_tkd, r_2014_tkd, r_2014_kdt} properties.
The ground-state magnetization was taken along $z$ axis and
the selfconsistent XC potentials were obtained in the local
spin-density approximation (LSDA) with parametrization
according to Ref.~\onlinecite{r_1980_vwn}.
The valence basis comprised $s$-, $p$-, and $d$-type orbitals and
the energy arguments for the propagators ${\bar g}_\pm$ and the
CPA-vertex corrections were obtained by adding a tiny imaginary
part $\pm \varepsilon$ to the real Fermi energy.
We have found that the dependence of the Gilbert damping parameter
on $\varepsilon$ is quite smooth and that the value of $\varepsilon
 = 10^{-6}$ Ry is sufficient for the studied systems, see
Fig.~\ref{f_alpeps} for an illustration.
Similar smooth dependences have been obtained also for other 
investigated alloys, such as Permalloy doped by $5d$ elements,
Heusler alloys, and stoichiometric FePt alloys with a partial
atomic long-range order.
In all studied cases, the number $N$ of ${\bf k}$ vectors needed
for reliable averaging over the Brillouin zone (BZ) was properly
checked; as a rule, $N \sim 10^8$ in the full BZ was sufficient
for most systems, but for diluted alloys (a few percent of 
impurities), $N \sim 10^9$ had to be taken. 

\begin{figure}
\altcn{ \includegraphics[width=0.55\textwidth]{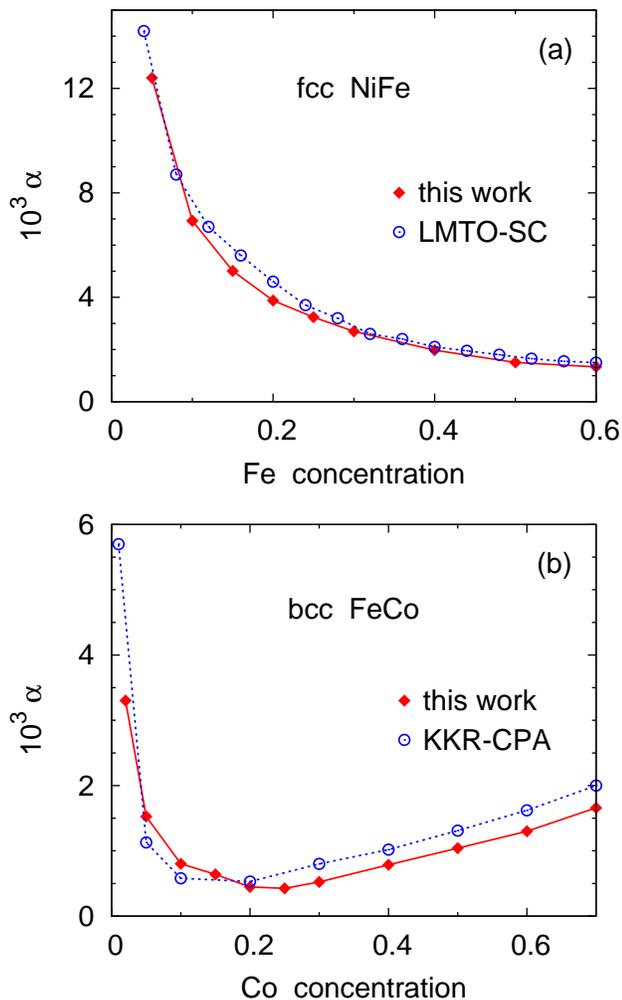} }%
{ \includegraphics[width=0.95\columnwidth]{fig_s3} }
\caption{(Color online)
The calculated concentration dependences of the Gilbert damping
parameter $\alpha$ for random fcc NiFe (a) and bcc FeCo (b) alloys.
The results of this work are marked by the full diamonds, whereas
the open circles depict the results of other approaches: the LMTO
supercell (LMTO-SC) technique \cite{r_2010_skb} and the KKR-CPA
method. \cite{r_2013_mkw}
\label{f_conc}}
\end{figure}

\subsection{Binary fcc and bcc solid solutions\label{ss_bss}}

The developed theory has been applied to random binary alloys
of $3d$ transition elements Fe, Co, and Ni, namely, to
the fcc NiFe and bcc FeCo alloys.
The most important results, including a comparison to other
existing \emph{ab initio} techniques, are summarized in
Fig.~\ref{f_conc}. 
One can see a good agreement of the calculated concentration
trends of the Gilbert damping parameter $\alpha = \alpha_{xx}
= \alpha_{yy}$ with the results of an LMTO-supercell approach
\cite{r_2010_skb} and of the KKR-CPA method. \cite{r_2013_mkw}
The decrease of $\alpha$ with increasing Fe content in the
concentrated NiFe alloys can be related to the increasing
alloy magnetization \cite{r_2010_skb} and to the decreasing
strength of the SO-interaction, \cite{r_2012_as} whereas the
behavior in the dilute limit can be explained by intraband
scattering due to Fe impurities. \cite{r_2007_gis, r_2007_vk,
r_2008_gis}
In the case of the FeCo system, the minimum of $\alpha$ around
20\% Co, which is also observed in room-temperature experiments,
\cite{r_2006_owy, r_1995_spf} is related primarily to a similar
concentration trend of the density of states at the Fermi
energy, \cite{r_2013_mkw} though the maximum of the magnetization
at roughly the same alloy composition \cite{r_1994_tkd} might
partly contribute as well.

\begin{table}
\caption{Comparison of the Gilbert damping parameter $\alpha$
for the fcc Ni$_{80}$Fe$_{20}$ random alloy (Permalloy)
calculated by the present approach and by other techniques
using the CPA or supercells (SC).
The last column displays the coherent part $\alpha^{\rm coh}$
of the total damping parameter according to 
Eq.~(\ref{eq_talpha_coh}).
The experimental value corresponds to room temperature.
\label{t_alppm}}
\begin{ruledtabular}
\begin{tabular}{lcc}
Method & $\alpha$ & $\alpha^{\rm coh}$ \\
\hline
This work, $\varepsilon = 10^{-5}$ Ry & $4.9\times 10^{-3}$ & 1.76 \\
This work, $\varepsilon = 10^{-6}$ Ry & $3.9\times 10^{-3}$ & 1.76 \\
KKR-CPA\footnote{Reference \onlinecite{r_2013_mkw}.} & 
                                        $4.2\times 10^{-3}$ & \\
LMTO-CPA\footnote{Reference \onlinecite{r_2012_as}.} & 
                                        $3.5\times 10^{-3}$ & \\
LMTO-SC\footnote{Reference \onlinecite{r_2010_skb}.} & 
                                        $4.6\times 10^{-3}$ & \\
Experiment\footnote{Reference \onlinecite{r_2006_owy}.} & 
                                        $8\times 10^{-3}$ & \\
\end{tabular}
\end{ruledtabular}
\end{table}

A more detailed comparison of all \emph{ab initio} results is
presented in Table~\ref{t_alppm} for the fcc Ni$_{80}$Fe$_{20}$
random alloy (Permalloy).
The differences in the values of $\alpha$ from the different
techniques can be ascribed to various theoretical features and
numerical details employed, such as the simplified treatment of
the SO-interaction in Ref.~\onlinecite{r_2012_as} instead of the
fully relativistic description, or the use of supercells in
Ref.~\onlinecite{r_2010_skb} instead of the CPA.
Taking into account that calculated residual resistivities for
this alloy span a wide interval between 2~$\mu\Omega$cm, 
see Ref.~\onlinecite{r_1997_bev, r_2012_tkd},
and 3.5~$\mu\Omega$cm, see Ref.~\onlinecite{r_2010_skb},
one can consider the scatter of the calculated values of $\alpha$
in Table~\ref{t_alppm} as little important.
The theoretical values of $\alpha$ are smaller systematically than
the measured values, typically by a factor of two.
This discrepancy might be partly due to the effects of finite
temperatures as well as due to additional structural defects of
real samples. 

A closer look at the theoretical results reveals that the total
damping parameters $\alpha$ are appreciably smaller than the
magnitudes of their coherent and vertex parts, see 
Table~\ref{t_alppm} for the case of Permalloy.
This is in agreement with the results of the model study in
Section \ref{ss_tcf}; similar conclusions about the importance
of the vertex corrections have been done with the XC-induced
torques in other CPA-based studies.  \cite{r_2011_emk, r_2013_mkw,
r_2015_as_a}
The present results prove that this unpleasant feature of the
nonlocal torques does not represent a serious obstacle in
obtaining reliable values of the Gilbert damping parameter
in random alloys.
We note that the vertex corrections can be negligible in approaches
employing the SO-induced torques, at least for systems with the
SO splittings much weaker than the XC splittings, \cite{r_2007_vk}
such as the binary ferromagnetic alloys of $3d$ transition metals,
\cite{r_2015_as_a} see also Section \ref{ss_tcf}.

\subsection{Pure iron with a model disorder \label{ss_fe}}

\begin{figure}
\altcn{ \includegraphics[width=0.55\textwidth]{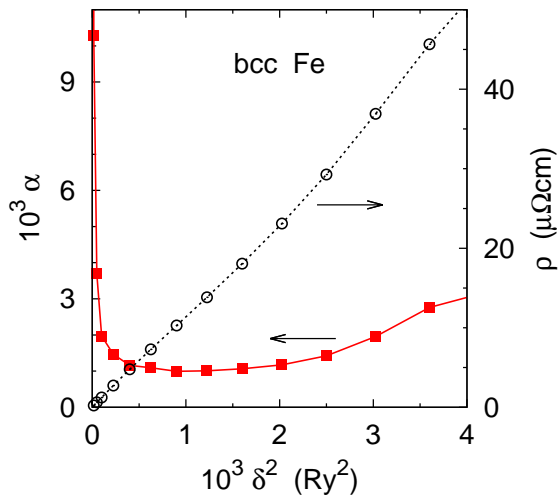} }%
{ \includegraphics[width=0.85\columnwidth]{fig_s4} }
\caption{(Color online)
The calculated Gilbert damping parameter $\alpha$ (full squares)
and the residual resistivity $\rho$ (open circles) of pure bcc
iron as functions of $\delta^2$, where $\delta$ is the strength
of a model atomic-level disorder. 
\label{f_fe}}
\end{figure}

As it has been mentioned in Section~\ref{s_intr}, the Gilbert
damping of pure ferromagnetic metals exhibits non-trivial
temperature dependences, which have been reproduced by
means of \emph{ab initio} techniques with various levels of
sophistication.
\cite{r_2007_gis, r_2007_vk, r_2011_lsy, r_2015_emc}
In this study, we have simulated the effect of finite temperatures
by introducing static fluctuations of the one-particle potential.
The adopted model of atomic-level disorder assumes that random
spin-independent shifts $\pm \delta$, constant inside each atomic
sphere and occurring with probabilities 50\% of both signs, are
added to the nonrandom selfconsistent potential obtained at zero
temperature.
The Fermi energy is kept frozen, equal to its selfconsistent
zero-temperature value.
This model can be easily treated in the CPA; the resulting Gilbert
damping parameter $\alpha$ of pure bcc Fe as a function of the
potential shift $\delta$ is plotted in Fig.~\ref{f_fe}.
 
The calculated dependence $\alpha(\delta)$ is nonmonotonic, with
a minimum at $\delta \approx 30$~mRy.
This trend is in a qualitative agreement with trends reported
previously by other authors, who employed phenomenological models
of the electron lifetime \cite{r_2007_gis, r_2007_vk} as well as
models for phonons and magnons. \cite{r_2011_lsy, r_2015_emc}
The origin of the nonmonotonic dependence $\alpha(\delta)$ has
been identified on the basis of the band structure of the
ferromagnetic system as an interplay between the intraband
contributions to $\alpha$, dominating for small values of $\delta$,
and the interband contributions, dominating for large values of
$\delta$. \cite{r_2004_sjl, r_2007_gis, r_2007_vk}
Since the present CPA-based approach does not use any bands, we
cannot perform a similar analysis.

The obtained minimum value of the Gilbert damping, 
$\alpha_{\rm min} \approx 10^{-3}$ (Fig.~\ref{f_fe}), agrees
reasonably well with the values obtained by the authors of
Ref.~\onlinecite{r_2007_gis, r_2007_vk, r_2011_lsy, r_2015_emc}.
This agreement indicates that the atomic-level disorder employed
here is equivalent to a phenomenological lifetime broadening.
For a rough quantitative estimation of the temperature effect,
one can employ the calculated resistivity $\rho$ of the model,
which increases essentially linearly with $\delta^2$, see
Fig.~\ref{f_fe}.
Since the metallic resistivity due to phonons increases linearly
with the temperature $T$ (for temperatures not much smaller than
the Debye temperature), one can assume a proportionality between 
$\delta^2$ and $T$.
The resistivity of bcc iron at the Curie temperature $T_{\rm C} =
1044$ K due to lattice vibrations can be estimated around
35 $\mu\Omega$cm, \cite{r_2015_emc, r_1959_wm} which sets an
approximate temperature scale to the data plotted in
Fig.~\ref{f_fe}.
However, a more accurate description of the temperature
dependence of the Gilbert damping parameter cannot be obtained,
mainly due to the neglected true atomic displacements and the
noncollinearity of magnetic moments (magnons). \cite{r_2015_emc} 

\subsection{FePt alloys with a partial long-range
            order \label{ss_fept}}

\begin{figure}
\altcn{ \includegraphics[width=0.55\textwidth]{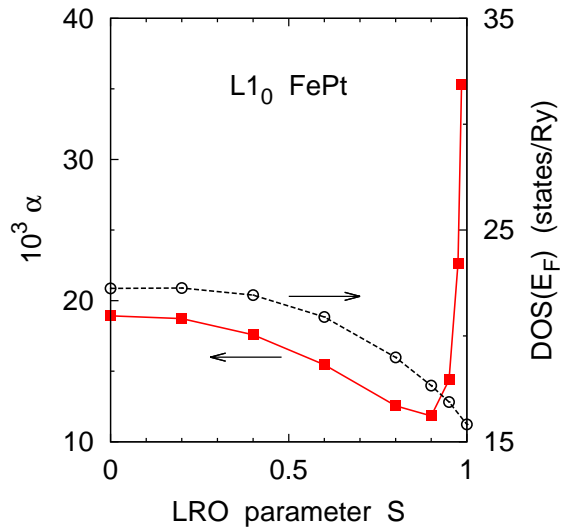} }%
{ \includegraphics[width=0.85\columnwidth]{fig_s5} }
\caption{(Color online)
The calculated Gilbert damping parameter $\alpha$ (full squares)
and the total DOS (per formula unit) at the Fermi energy (open
circles) of stoichiometric L1$_0$ FePt alloys as functions of the
LRO parameter $S$. 
\label{f_fept}}
\end{figure}

Since important ferromagnetic materials include ordered alloys,
we address here the Gilbert damping in stoichiometric FePt alloys
with L1$_0$ atomic long-range order (LRO).
Their transport properties \cite{r_2014_kdt} and the damping
parameter \cite{r_2012_as} have recently been studied by means of
the TB-LMTO method in dependence on a varying degree of the LRO. 
These fcc-based systems contain two sublattices with respective
occupations Fe$_{1-y}$Pt$_y$ and Pt$_{1-y}$Fe$_y$
where $y$ ($0 \leq y \leq 0.5$) denotes the concentration
of antisite atoms.
The LRO parameter $S$ ($0 \leq S \leq 1$) is then defined as
$S = 1 - 2y$, so that $S = 0$ corresponds to the random
fcc alloy and $S = 1$ corresponds to the perfectly ordered
L1$_0$ structure.

The resulting Gilbert damping parameter is displayed in
Fig.~\ref{f_fept} as a function of $S$.
The obtained trend with a broad maximum at $S = 0$ and a 
minimum around $S = 0.9$ agrees very well with the previous
result. \cite{r_2012_as}
The values of $\alpha$ in Fig.~\ref{f_fept} are about 10\% higher
than those in Ref.~\onlinecite{r_2012_as}, which can be ascribed
to the fully relativistic treatment in the present study in
contrast to a simplified treatment of the SO interaction in
Ref.~\onlinecite{r_2012_as}.
The Gilbert damping in the FePt alloys is an order of magnitude
stronger than in the alloys of $3d$ elements (Section \ref{ss_bss})
owing to the stronger SO interaction of Pt atoms.
The origin of the slow decrease of $\alpha$ with increasing $S$
(for $0 \leq S \leq 0.9$) can be explained by the decreasing
total density of states (DOS) at the Fermi energy, see
Fig.~\ref{f_fept}, which represents an analogy to a similar
correlation observed, e.g., for bcc FeCo alloys. \cite{r_2013_mkw}

All calculated values of $\alpha$ shown in Fig.~\ref{f_fept},
corresponding to $0 \leq S \leq 0.985$, are appreciably smaller
than the measured one which amounts to $\alpha \approx 0.06$
reported for a thin L1$_0$ FePt epitaxial film. \cite{r_2011_mii} 
The high measured value of $\alpha$ might be thus explained by the
present calculations by assuming a very small concentration
of antisites in the prepared films, which does not seem too
realistic. 
Another potential source of the discrepancy lies in the thin-film
geometry used in the experiment.
Moreover, the divergence of $\alpha$ in the limit of $S \to 1$ 
(Fig.~\ref{f_fept}) illustrates a general shortcoming of approaches
based on the torque-correlation formula (\ref{eq_tcf}), since the
zero-temperature Gilbert damping parameter of a pure ferromagnet
should remain finite.
A correct treatment of this case, including the dilute limit of
random alloys (Fig.~\ref{f_conc}), must take into account the full
interacting susceptibility in the presence of SO
interaction. \cite{r_2009_gm_a, r_1972_kp} 
Pilot \emph{ab initio} studies in this direction have recently
appeared for nonrandom systems; \cite{r_2015_ssb, r_2015_lss}
however, their extension to disordered systems goes far beyond
the scope of this work.

\section{Conclusions\label{s_conc}}

We have introduced nonlocal torques as an alternative to the usual
local torque operators entering the torque-correlation formula
for the Gilbert damping tensor.
Within the relativistic TB-LMTO-ASA method, this idea leads to
effective nonlocal torques as non-site-diagonal and spin-independent
matrices.
For substitutionally disordered alloys, the nonlocal torques are
nonrandom, which allows one to develop an internally consistent
theory in the CPA.
The CPA-vertex corrections proved indispensable for an exact
equivalence of the nonlocal nonrandom torques with their local
random counterparts.
The concept of the nonlocal torques is not limited to the LMTO
method and its formulation both in a semiempirical TB theory
and in the KKR theory is straightforward.

The numerical implementation and the results for binary solid
solutions show that the total Gilbert damping parameters from
the nonlocal torques are much smaller than magnitudes of the
coherent parts and of the vertex corrections.
Nevertheless, the total damping parameters for the studied
NiFe, FeCo and FePt alloys compare quantitatively very well
with results of other \emph{ab initio} techniques, 
\cite{r_2010_skb, r_2012_as, r_2013_mkw}
which indicates a fair numerical stability of the developed
theory.

The performed numerical study of the Gilbert damping in pure bcc
iron as a function of an atomic-level disorder yields a
nonmonotonic dependence in a qualitative agreement with the trends
consisting of the conductivity-like and resistivity-like regions,
obtained from a phenomenological quasiparticle lifetime broadening
\cite{r_2004_sjl, r_2007_gis, r_2007_vk} or from the
temperature-induced frozen phonons \cite{r_2011_lsy, r_2013_mkw}
and magnons. \cite{r_2015_emc}
Future studies should clarify the applicability of the introduced
nonlocal torques to a full quantitative description of the
finite-temperature behavior as well as to other torque-related
phenomena, such as the spin-orbit torques due to applied electric
fields. \cite{r_2009_mz, r_2014_fbm}

%

\begin{acknowledgments}
The authors acknowledge financial support by the Czech Science
Foundation (Grant No.\ 15-13436S).
\end{acknowledgments}


\appendix
\section{Torque correlation formula in a matrix
          representation\label{app_matrep}}

In this Appendix, evaluation of the Kubo-Greenwood expression
for the torque-correlation formula (\ref{eq_tcf}) is discussed
in the case of the XC-induced torque operators using matrix
representations of all operators in an orthonormal basis that
varies due to the varying direction of the XC field.
All operators are denoted by a hat, in order to be distinguished
from matrices representing these operators in the chosen basis.
Let us consider a one-particle Hamiltonian ${\hat H} = {\hat H} 
(\theta_1, \theta_2)$ depending on two real variables $\theta_j$,
$j=1,2$, and let us denote ${\hat T}^{(j)} (\theta_1, \theta_2) = 
\partial {\hat H}(\theta_1, \theta_2) / \partial \theta_j$.
In our case, the variables $\theta_j$ play the role of rotation
angles and the operators ${\hat T}^{(j)}$ are the corresponding
torques.
Let us denote the resolvents of ${\hat H} (\theta_1, \theta_2)$
at the Fermi energy as ${\hat G}_\pm (\theta_1, \theta_2)$ and
let us consider a special linear response coefficient (arguments
$\theta_1$ and $\theta_2$ are omitted here and below for brevity) 
\begin{eqnarray}
\label{eq_app_alfadef}
c & = & {\rm Tr} \{
{\hat T}^{(1)} ( {\hat G}_+ - {\hat G}_- )
{\hat T}^{(2)} ( {\hat G}_+ - {\hat G}_- )
\} 
\altcn{ \nonumber\\ }{ \\ }
 & = & {\rm Tr} \{
(\partial {\hat H} / \partial \theta_1) ( {\hat G}_+ - {\hat G}_- )
(\partial {\hat H} / \partial \theta_2) ( {\hat G}_+ - {\hat G}_- )
\} .
\altcn{ }{ \nonumber }
\end{eqnarray}
This torque-correlation coefficient equals the Gilbert damping
parameter (\ref{eq_tcf}) with the prefactor $(-\alpha_0)$ suppressed.
For its evaluation, we introduce an orthonormal basis
$| \chi_m (\theta_1, \theta_2) \rangle$ and represent all operators
in this basis.
This leads to matrices 
$H(\theta_1, \theta_2) = \{ H_{mn} (\theta_1, \theta_2) \}$,
$G_\pm(\theta_1, \theta_2) = \{ (G_\pm)_{mn} (\theta_1, \theta_2) \}$
and $T^{(j)} (\theta_1, \theta_2) = \{ T^{(j)}_{mn} (\theta_1,
 \theta_2) \}$, where
\begin{eqnarray}
H_{mn} & = & \langle \chi_m | {\hat H} | \chi_n \rangle , \qquad
(G_\pm)_{mn} = \langle \chi_m | {\hat G}_\pm | \chi_n \rangle , 
\nonumber\\
T^{(j)}_{mn} & = & \langle \chi_m | {\hat T}^{(j)} | \chi_n \rangle 
 = \langle \chi_m | \partial {\hat H} / \partial \theta_j | 
\chi_n \rangle , 
\label{eq_app_hgtmat}
\end{eqnarray}
and, consequently, to the response coefficient (\ref{eq_app_alfadef})
expressed by using the matrices (\ref{eq_app_hgtmat}) as
\begin{equation}
c = {\rm Tr} \{
T^{(1)} ( G_+ - G_- ) T^{(2)} ( G_+ - G_- ) \} .
\label{eq_app_alfamat}
\end{equation}
However, in evaluation of the last expression, attention has to be
paid to the difference between the matrix $T^{(j)} (\theta_1, 
\theta_2)$ and the partial derivative of the matrix 
$H (\theta_1, \theta_2)$ with respect to $\theta_j$.
This difference follows from the identity ${\hat H} = \sum_{mn} 
| \chi_m \rangle H_{mn} \langle \chi_n |$, which yields
\altcn{ \begin{equation} }{ \begin{eqnarray} }
T^{(j)}_{mn} 
\altcn{ = }{ & = & }
\partial H_{mn} / \partial \theta_j 
+ \sum_k \langle \chi_m | \partial \chi_k / \partial \theta_j 
\rangle H_{kn}
\altcn{ + }{ \nonumber\\ & & + }
\sum_k H_{mk} \langle \partial \chi_k / \partial \theta_j
| \chi_n \rangle ,
\label{eq_app_met}
\altcn{ \end{equation} }{ \end{eqnarray} }
where we employed the orthogonality relations $\langle \chi_m 
(\theta_1, \theta_2) | \chi_n (\theta_1, \theta_2) \rangle = 
\delta_{mn}$.
Their partial derivatives yield 
\begin{equation}
\langle \chi_m | \partial \chi_n / \partial \theta_j \rangle = 
- \langle \partial \chi_m / \partial \theta_j | \chi_n \rangle
\equiv Q^{(j)}_{mn} ,
\label{eq_app_meq}
\end{equation}
where we introduced elements of matrices 
$Q^{(j)} = \{ Q^{(j)}_{mn} \}$ for $j=1,2$.
Note that the matrices $Q^{(j)} (\theta_1, \theta_2)$ reflect
explicitly the dependence of the basis vectors $| \chi_m 
(\theta_1, \theta_2) \rangle$ on $\theta_1$ and $\theta_2$.
The relation (\ref{eq_app_met}) between the matrices $T^{(j)}$ and
$\partial H / \partial \theta_j$ can be now rewritten compactly as
\begin{equation}
T^{(j)} = \partial H / \partial \theta_j + [ Q^{(j)} , H ] .
\label{eq_app_mrt}
\end{equation}
Since the last term has a form of a commutator with the Hamiltonian
matrix $H$, the use of Eq.~(\ref{eq_app_mrt}) in the formula
(\ref{eq_app_alfamat}) leads to the final matrix expression for the
torque correlation,
\begin{equation}
c = {\rm Tr} \{
(\partial H / \partial \theta_1) ( G_+ - G_- ) 
(\partial H / \partial \theta_2) ( G_+ - G_- ) \} .
\label{eq_app_alfafin}
\end{equation}
The equivalence of Eqs.~(\ref{eq_app_alfamat}) and 
(\ref{eq_app_alfafin}) rests on the rules $[ Q^{(j)} , H ] = 
[ E_{\rm F} - H , Q^{(j)} ]$ and $( E_{\rm F} - H ) ( G_+ - G_- )
= ( G_+ - G_- ) ( E_{\rm F} - H ) = 0$ and on the cyclic
invariance of the trace.
It is also required that the matrices $Q^{(j)}$ are compatible with
periodic boundary conditions used in calculations of extended
systems, which is obviously the case for angular variables $\theta_j$
related to the global changes (uniform rotations) of the
magnetization direction.

The obtained result means that the original response coefficient
(\ref{eq_app_alfadef}) involving the torques as angular derivatives
of the Hamiltonian can be expressed solely by using matrix elements
of the Hamiltonian in an angle-dependent basis; the angular
dependence of the basis vectors does not enter explicitly the final
torque-correlation formula (\ref{eq_app_alfafin}).

\section{LMTO Hamiltonian of a ferromagnet with a tilted
         magnetic field\label{app_tilt}}

Here we sketch a derivation of the fully relativistic LMTO
Hamiltonian matrix for a ferromagnet with the XC-field direction
tilted from a reference direction along an easy axis.
The derivation rests on the form of the Kohn-Sham-Dirac Hamiltonian
in the LMTO-ASA method. \cite{r_1991_slg, r_1996_sdk, r_1997_tdk}
The symbols with superscript $0$ refer to the reference system,
the symbols without this superscript refer to the system with the
tilted XC field. 
The operators (Hamiltonians, rotation operators) are denoted by
symbols with a hat. 
The spin-dependent parts of the ASA potentials due to the 
XC fields are rigidly rotated while the spin-independent parts
are unchanged, in full analogy to the approach employed in
the relativistic KKR method. \cite{r_2011_emk, r_2013_mkw}

The ASA-Hamiltonians of both systems are given by lattice sums 
${\hat H}^0 = \sum_{\bf R} {\hat H}^0_{\bf R}$ and
${\hat H} = \sum_{\bf R} {\hat H}_{\bf R}$,
where the individual site-contributions are coupled mutually by
${\hat H}_{\bf R} = {\hat U}_{\bf R} {\hat H}^0_{\bf R}
{\hat U}^+_{\bf R}$, where ${\hat U}_{\bf R}$ denotes the unitary
operator of a rotation (in the orbital and spin space) around the
${\bf R}$th lattice site which brings the local XC field from its
reference direction into the tilted one.
Let $| \phi^0_{{\bf R}\Lambda} \rangle$ and
$| {\dot \phi}^0_{{\bf R}\Lambda} \rangle$ denote, respectively,
the phi and phi-dot orbitals of the reference Hamiltonian
${\hat H}^0_{\bf R}$, then
\begin{equation}
| \phi_{{\bf R}\Lambda} \rangle = 
{\hat U}_{\bf R} | \phi^0_{{\bf R}\Lambda} \rangle,
\qquad
| {\dot \phi}_{{\bf R}\Lambda} \rangle = 
{\hat U}_{\bf R} | {\dot \phi}^0_{{\bf R}\Lambda} \rangle
\label{eq_app_phphd}
\end{equation}
define the phi and phi-dot orbitals of the Hamiltonian
${\hat H}_{\bf R}$.
The orbital index $\Lambda$ labels all linearly independent
solutions (regular at the origin) of the spin-polarized relativistic
single-site problem; the detailed structure of $\Lambda$ can be
found elsewhere. \cite{r_1991_slg, r_1996_sdk, r_1997_tdk}
Let us introduce further the well-known empty-space solutions
$| K^{\infty,0}_{{\bf R}N} \rangle$ (extending over the whole real
space), $| K^{\mathrm{int},0}_{{\bf R}N} \rangle$ (extending over
the interstitial region), and $| K^0_{{\bf R}N} \rangle$ and 
$| J^0_{{\bf R}N} \rangle$ (both truncated outside the ${\bf R}$th
sphere), needed for the definition of the LMTOs of the reference
system. \cite{r_1984_aj, r_1985_ajg, r_1986_apj}
Their index $N$, which defines the spin-spherical harmonics of the
large component of each solution, can be taken either in the
nonrelativistic $(\ell m s)$ form or in its relativistic 
$(\kappa \mu)$ counterpart.
We define further 
\begin{equation}
| Z_{{\bf R}N} \rangle = {\hat U}_{\bf R} | Z^0_{{\bf R}N} \rangle 
\qquad \mathrm{for} \ \, 
Z = K^\infty , \, K , \, J .
\label{eq_app_rotz1}
\end{equation}
Isotropy of the empty space guarantees relations (for 
$Z = K^\infty$, $K$, $J$)
\altcn{ \begin{equation} }{ \begin{eqnarray} }
| Z_{{\bf R}N} \rangle 
\altcn{ = }{ & = & }
\sum_{N'} | Z^0_{{\bf R}N'} \rangle U_{N'N} ,
\altcn{ \qquad }{ \nonumber\\ }
| Z^0_{{\bf R}N} \rangle 
\altcn{ = }{ & = & }
\sum_{N'} | Z_{{\bf R}N'} \rangle U^+_{N'N} ,
\label{eq_app_rotz2}
\altcn{ \end{equation} }{ \end{eqnarray} }
where $U = \{ U_{N'N} \}$ denotes a unitary matrix representing
the rotation in the space of spin-spherical harmonics 
and where $U^+_{N'N} \equiv (U^+)_{N'N} = (U_{NN'})^\ast
= (U^{-1})_{N'N}$; the matrix $U$ is the same for all lattice sites
${\bf R}$ since we consider only uniform rotations of the XC-field
direction inside the ferromagnet.
The expansion theorem for the envelope orbital 
$| K^{\infty,0}_{{\bf R}N} \rangle$ is
\altcn{ \begin{equation} }{ \begin{eqnarray} }
| K^{\infty,0}_{{\bf R}N} \rangle 
\altcn{ = }{ & = & }
| K^{\mathrm{int},0}_{{\bf R}N} \rangle +
| K^0_{{\bf R}N} \rangle 
\altcn{ - }{ \nonumber\\ & & - }
\sum_{{\bf R}'N'} | J^0_{{\bf R}'N'} \rangle
S^0_{{\bf R}'N',{\bf R}N} ,
\label{eq_app_exp0}
\altcn{ \end{equation} }{ \end{eqnarray} }
where $S^0_{{\bf R}'N',{\bf R}N}$ denote elements of the canonical
structure-constant matrix (with vanishing on-site elements, 
$S^0_{{\bf R}N',{\bf R}N} = 0$) of the reference system.
The use of relations (\ref{eq_app_rotz2}) in the expansion
(\ref{eq_app_exp0}) together with an abbreviation
\begin{equation}
| K^\mathrm{int}_{{\bf R}N} \rangle = \sum_{N'} 
| K^{\mathrm{int},0}_{{\bf R}N'} \rangle U_{N'N} 
\label{eq_app_kint}
\end{equation}
yields the expansion of the envelope orbital 
$| K^\infty_{{\bf R}N} \rangle$ as 
\altcn{ \begin{equation} }{ \begin{eqnarray} }
| K^\infty_{{\bf R}N} \rangle 
\altcn{ = }{ & = & } 
| K^\mathrm{int}_{{\bf R}N} \rangle + 
| K_{{\bf R}N} \rangle 
\altcn{ - }{ \nonumber\\ & & - }
\sum_{{\bf R}'N'} | J_{{\bf R}'N'} \rangle
( U^+ S^0 U )_{{\bf R}'N',{\bf R}N} ,
\label{eq_app_exp2}
\altcn{ \end{equation} }{ \end{eqnarray} }
where $U$ and $U^+$ denote site-diagonal matrices with elements
$U_{{\bf R}'N',{\bf R}N} = \delta_{{\bf R}'{\bf R}} U_{N'N}$ and
$(U^+)_{{\bf R}'N',{\bf R}N} = \delta_{{\bf R}'{\bf R}} U^+_{N'N}$.
Note the same form of expansions (\ref{eq_app_exp0}) and
(\ref{eq_app_exp2}), with the orbitals $| Z^0_{{\bf R}N} \rangle$
replaced by the rotated orbitals $| Z_{{\bf R}N} \rangle$ 
($Z = K^\infty , K , J$), with the interstitial parts 
$| K^{\mathrm{int},0}_{{\bf R}N} \rangle$ replaced by their linear
combinations $| K^\mathrm{int}_{{\bf R}N} \rangle$, and with the 
structure-constant matrix $S^0$ replaced by the product $U^+ S^0 U$.

The non-orthogonal LMTO $| \chi^0_{{\bf R}N} \rangle$ for the
reference system is obtained from the expansion (\ref{eq_app_exp0}),
in which all orbitals $| K^0_{{\bf R}N} \rangle$ and 
$| J^0_{{\bf R}N} \rangle$ are replaced by linear combinations of 
$| \phi^0_{{\bf R}\Lambda} \rangle$ and
$| {\dot \phi}^0_{{\bf R}\Lambda} \rangle$.
A similar replacement of the orbitals 
$| K_{{\bf R}N} \rangle$ and $| J_{{\bf R}N} \rangle$ by linear
combinations of $| \phi_{{\bf R}\Lambda} \rangle$ and
$| {\dot \phi}_{{\bf R}\Lambda} \rangle$ in the expansion
(\ref{eq_app_exp2}) yields the non-orthogonal LMTO 
$| \chi_{{\bf R}N} \rangle$ for the system with the tilted XC field.
The coefficients in these linear combinations---obtained from 
conditions of continuous matching at the sphere boundaries and
leading directly to the LMTO potential parameters---are identical
for both systems, as follows from the rotation relations 
(\ref{eq_app_phphd}) and (\ref{eq_app_rotz1}).
For these reasons, the only essential difference between both systems
in the construction of the non-orthogonal and orthogonal LMTOs (and
of the accompanying Hamiltonian and overlap matrices in the ASA) is
due to the difference between the matrices $S^0$ and $U^+ S^0 U$.

As a consequence, the LMTO Hamiltonian matrix in the orthogonal LMTO
representation for the system with a tilted magnetization is easily
obtained from that for the reference system, Eq.~(\ref{eq_hlmto}),
and it is given by
\begin{equation}
H = C + (\sqrt{\Delta})^+ U^+ S U
( 1 - \gamma U^+ S U )^{-1} \sqrt{\Delta} ,
\label{eq_app_hortf}
\end{equation}
where the $C$, $\sqrt{\Delta}$ and $\gamma$ are site-diagonal
matrices of the potential parameters of the reference system and
where we suppressed the superscript $0$ at the
structure-constant matrix $S$ of the reference system.
Note that the dependence of $H$ on the XC-field direction is
contained only in the similarity transformation $U^+ S U$ of the
original structure-constant matrix $S$ generated by the rotation
matrix $U$.
For the rotation by an angle $\theta$ around an axis along a unit
vector ${\bf n}$, the rotation matrix is given by 
$U(\theta) = \exp(- {\rm i} {\bf n} \cdot {\bf J} \theta)$,
where the site-diagonal matrices ${\bf J} \equiv 
( {\cal J}^x , {\cal J}^y , {\cal J}^z )$ with matrix elements 
${\cal J}^\mu_{{\bf R}'N',{\bf R}N} = \delta_{{\bf R}'{\bf R}} 
{\cal J}^\mu_{N'N}$ ($\mu = x , y , z$) reduce to usual matrices
of the total (orbital plus spin) angular momentum operator.
The limit of small $\theta$ yields $U(\theta) \approx 1 - {\rm i} 
{\bf n} \cdot {\bf J} \theta$, which leads to the
$\theta$-derivative of the Hamiltonian matrix (\ref{eq_app_hortf})
at $\theta = 0$:
\begin{equation}
\partial H / \partial \theta = 
{\rm i} ( F^+ )^{-1} [ {\bf n} \cdot {\bf J} , S ] F^{-1} ,
\label{eq_app_dhdt}
\end{equation}
where we abbreviated $F = ( \sqrt{\Delta} )^{-1} (1 - \gamma S)$ 
and $F^+ = (1 - S \gamma) [ (\sqrt{\Delta})^+ ]^{-1}$.
Since the structure-constant matrix $S$ is spin-independent,
the total angular momentum operator ${\bf J}$ in
(\ref{eq_app_dhdt}) can be replaced by its orbital momentum
counterpart ${\bf L} \equiv 
( {\cal L}^x , {\cal L}^y , {\cal L}^z )$, so that
\begin{equation}
\partial H / \partial \theta =
{\rm i} ( F^+ )^{-1} [ {\bf n} \cdot {\bf L} , S ] F^{-1} .
\label{eq_app_dhdto}
\end{equation}
The relations (\ref{eq_app_dhdt}) and (\ref{eq_app_dhdto}) are used
to derive the LMTO-ASA torque-correlation formula (\ref{eq_talpha}).

\section{Equivalence of the Gilbert damping in the CPA
          with local and nonlocal torques (Supplemental Material)}

\subsection{Introductory remarks}

The problem of equivalence of the Gilbert damping tensor expressed
with the local (loc) and nonlocal (nl) torques can be reduced to
the problem of equivalence of these two expressions:
\begin{eqnarray}
\alpha^{\rm loc} & = & \alpha_0 \, {\rm Tr} \langle
( g_+ - g_- ) [ P , K ] ( g_+ - g_- ) [ P , K ] \rangle
\nonumber\\
 & = & \alpha_0 \sum_{p = \pm} \sum_{q = \pm} {\rm sgn} (pq)
\, {\rm Tr} \langle g_p [ P , K ] g_q [ P , K ] \rangle 
\nonumber\\
 & = & \alpha_0 \sum_{p = \pm} \sum_{q = \pm} {\rm sgn} (pq)
\beta^{\rm loc}_{pq} ,
\label{sm_alpha_loc}
\end{eqnarray}
and
\begin{eqnarray}
\alpha^{\rm nl} & = & \alpha_0 \, {\rm Tr} \langle
( g_+ - g_- ) [ K , S ] ( g_+ - g_- ) [ K , S ] \rangle
\nonumber\\
 & = & \alpha_0 \sum_{p = \pm} \sum_{q = \pm} {\rm sgn} (pq)
\, {\rm Tr} \langle g_p [ K , S ] g_q [ K , S ] \rangle 
\nonumber\\
 & = & \alpha_0 \sum_{p = \pm} \sum_{q = \pm} {\rm sgn} (pq)
\beta^{\rm nl}_{pq} .
\label{sm_alpha_nl}
\end{eqnarray}
The symbols ${\rm Tr}$ and $\langle \dots \rangle$ and the
quantities $\alpha_0$, $g_\pm$, $P$ and $S$ have the same meaning
as in the main text and the quantity $K$ substitutes any of the
operators (matrices) ${\cal J}^\mu$ or ${\cal L}^\mu$.
Note that owing to the symmetric nature of the original damping
tensors, the analysis can be confined to scalar quantities
$\alpha^{\rm loc}$ and $\alpha^{\rm nl}$ depending on a general
site-diagonal nonrandom operator $K$.
The choice of $K = K^\mu$ in (\ref{sm_alpha_loc}) and 
(\ref{sm_alpha_nl}) produces the diagonal elements of both
tensors, whereas the choice of $K = K^\mu \pm K^\nu$ for
$\mu \neq \nu$ leads to all off-diagonal elements.
The quantities $\beta^{\rm loc}_{pq}$ and $\beta^{\rm nl}_{pq}$
are expressions of the form
\begin{eqnarray}
\beta^{\rm loc} & = & {\rm Tr} \langle g^1 ( P^1 K - K P^2 )
g^2 ( P^2 K - K P^1 ) \rangle ,
\nonumber\\
\beta^{\rm nl} & = & {\rm Tr} \langle g^1 [ K , S ]
g^2 [ K , S ] \rangle ,
\label{sm_beta_def}
\end{eqnarray}
where the $g^1$ and $g^2$ replace the $g_p$ and $g_q$, respectively.
For an internal consistency of these and following expressions,
we have also introduced $P^1 = P^2 = P$.

This supplement contains a proof of the equivalence of
$\beta^{\rm loc}$ and $\beta^{\rm nl}$ and, consequently, of
$\alpha^{\rm loc}$ and $\alpha^{\rm nl}$. 
The CPA-average in $\beta^{\rm nl}$ with a nonlocal nonrandom
torque has been done using the theory by Velick\'y \cite{r_1969_bv}
as worked out in detail within the present LMTO formalism by Carva
et al.~\cite{r_2006_ctk} whereas the averaging in $\beta^{\rm loc}$
involving a local but random torque has been treated using the
approach by Butler. \cite{r_1985_whb}

\subsection{Auxiliary quantities and relations} 

Since the necessary formulas of the CPA in multiorbital techniques
\cite{r_2006_ctk, r_1985_whb} are little transparent, partly owing
to the complicated indices of two-particle quantities, we employ
here a formalism with the lattice-site index ${\bf R}$ kept but
with all orbital indices suppressed.

The Hilbert space is a sum of mutually orthogonal subspaces of
individual lattice sites ${\bf R}$; the corresponding projectors
will be denoted by $\Pi_{\bf R}$.
A number of relevant operators are site-diagonal, i.e., they can be
written as $X = \sum_{\bf R} X_{\bf R}$, where the site contributions
are given by $X_{\bf R} = \Pi_{\bf R} X = X \Pi_{\bf R} = 
\Pi_{\bf R} X \Pi_{\bf R}$.
Such operators are, e.g., the random potential functions,
$P^j = \sum_{\bf R} P^j_{\bf R}$, and the nonrandom coherent
potential functions ${\cal P}^j = \sum_{\bf R} {\cal P}^j_{\bf R}$,
where $j=1,2$.
The operator $K$ in (\ref{sm_beta_def}) is site-diagonal as well,
but its site contributions $K_{\bf R}$ will not be used explicitly
in the following.

Among the number of CPA-relations for single-particle properties,
we will use the equation of motion for the average auxiliary
Green's functions ${\bar g}^j$ ($j=1,2$), 
\begin{equation}
{\bar g}^j ( {\cal P}^j - S ) = ( {\cal P}^j - S ) {\bar g}^j = 1 ,
\label{sm_eom}
\end{equation}
as well as the definition of random single-site t-matrices
$t^j_{\bf R}$ ($j=1,2$) with respect to the effective CPA-medium,
given by
\begin{equation}
t^j_{\bf R} = ( P^j_{\bf R} - {\cal P}^j_{\bf R} ) 
[ 1 + {\bar g}^j ( P^j_{\bf R} - {\cal P}^j_{\bf R} ) ]^{-1} .
\label{sm_sstm}
\end{equation}
The operators $t^j_{\bf R}$ are site-diagonal, being non-zero only
in the subspace of site ${\bf R}$. 
The last definition leads to identities
\begin{eqnarray}
( 1 - \tmo \bgo ) \pfo & = & \cpo + \tmo ( 1 - \bgo \cpo ) ,
\nonumber\\
\pft ( 1 - \bgt \tmt ) & = & \cpt + ( 1 - \cpt \bgt ) \tmt ,
\label{sm_tmid}
\end{eqnarray}
which will be employed below together with the CPA-selfconsistency
conditions $\langle t^j_{\bf R} \rangle = 0$ ($j=1,2$).

For the purpose of evaluation of the two-particle averages in
(\ref{sm_beta_def}), we introduce several nonrandom operators:
\begin{equation}
f^{12} = \bgo K - K \bgt , 
\qquad
\zeta^{12} = \bgo [ K , S ] \bgt ,
\label{sm_f_zeta_def}
\end{equation}
and a site-diagonal operator $\gamma^{12} = \sum_{\bf R}
\gamma^{12}_{\bf R}$, where
\begin{equation}
\gamma^{12} = {\cal P}^1 K - K {\cal P}^2 ,
\qquad
\gamma^{12}_{\bf R} = \cpo K - K \cpt .
\label{sm_gamma_def}
\end{equation}
By interchanging the superscripts $1 \leftrightarrow 2$ in
(\ref{sm_f_zeta_def}) and (\ref{sm_gamma_def}), one can also
get quantities $f^{21}$, $\zeta^{21}$, $\gamma^{21}$ and
$\gamma^{21}_{\bf R}$; this will be implicitly understood in
the relations below as well.
The three operators $f^{12}$, $\zeta^{12}$ and $\gamma^{12}$
satisfy a relation
\begin{equation}
f^{12} + \zeta^{12} + \bgo \gamma^{12} \bgt = 0 ,
\label{sm_f_zeta_gamma}
\end{equation}
which can be easily proved from their definitions
(\ref{sm_f_zeta_def}) and (\ref{sm_gamma_def}) and from the
equation of motion (\ref{sm_eom}).
Another quantity to be used in the following is a nonrandom
site-diagonal operator $\vartheta^{12}$ related to the local torque
and defined by
\begin{eqnarray}
\vartheta^{12}_{\bf R} & = & \langle ( 1 - \tmo \bgo ) 
( \pfo K - K \pft ) ( 1 - \bgt \tmt ) \rangle ,
\nonumber\\
\vartheta^{12} & = & \sum_{\bf R} \vartheta^{12}_{\bf R} .
\label{sm_theta_def}
\end{eqnarray}
Its site contributions can be rewritten explicitly as
\begin{equation}
\vartheta^{12}_{\bf R} = \gamma^{12}_{\bf R} + \langle \tmo 
( f^{12} + \bgo \gamma^{12}_{\bf R} \bgt ) \tmt \rangle .
\label{sm_theta_gamma}
\end{equation}
The last relation follows from the definition (\ref{sm_theta_def}),
from the identities (\ref{sm_tmid}) and from the CPA-selfconsistency
conditions.
Moreover, the site contributions $\vartheta^{12}_{\bf R}$ and
$\gamma^{12}_{\bf R}$ satisfy a sum rule
\begin{equation}
\gamma^{12}_{\bf R} = 
{\sum_{{\bf R}'}}^\prime \langle \tmo \bgo \gamma^{12}_{{\bf R}'}
\bgt \tmt \rangle + \langle \tmo \zeta^{12} \tmt \rangle
+ \vartheta^{12}_{\bf R} ,
\label{sm_sumrule1}
\end{equation}
where the prime at the sum excludes the term with
${\bf R}' = {\bf R}$.
This sum rule can be proved by using the definitions of 
$\zeta^{12}$ (\ref{sm_f_zeta_def}) and $\gamma^{12}_{\bf R}$
(\ref{sm_gamma_def}) and by employing the previous relation
for $\vartheta^{12}_{\bf R}$ (\ref{sm_theta_gamma}) and the
equation of motion (\ref{sm_eom}).

The treatment of two-particle quantities requires the use of
a direct product $a \otimes b$ of two operators $a$ and $b$.
This is equivalent to the concept of a superoperator, i.e.,
a linear mapping defined on the vector space of all linear
operators.
In this supplement, superoperators are denoted by an overhat,
e.g., $\hat{m}$.
In the present formalism, the direct product of two operators
$a$ and $b$ can be identified with a superoperator
$\hat{m} = a \otimes b$, which induces a mapping
\begin{equation}
x \mapsto \hat{m} x = ( a \otimes b ) x = axb ,
\label{sm_dirprod}
\end{equation}
where $x$ denotes an arbitrary usual operator.
This definition leads, e.g., to a superoperator multiplication rule
\begin{equation}
( a \otimes b ) ( c \otimes d ) = ( a c ) \otimes ( d b ) .
\label{sm_multrule}
\end{equation}
In the CPA, the most important superoperators are
\begin{equation}
\hat{w}^{12} = \sum_{\bf R} \langle \tmo \otimes \tmt \rangle
\label{sm_w12_def}
\end{equation}
and
\begin{equation}
\hat{\chi}^{12} = {\sum_{{\bf R} {\bf R}'}}^\prime 
\Pi_{\bf R} \bgo \Pi_{{\bf R}'} \otimes 
\Pi_{{\bf R}'} \bgt \Pi_{\bf R} 
\label{sm_chi12_def}
\end{equation}
where the prime at the double sum excludes the terms with
${\bf R} = {\bf R}'$.
The quantity $\hat{w}^{12}$ represents the irreducible CPA-vertex
and the quantity $\hat{\chi}^{12}$ corresponds to a restricted
two-particle propagator with excluded on-site terms.
By using these superoperators, the previous sum rule
(\ref{sm_sumrule1}) can be rewritten compactly as
\begin{equation}
( \hat{1} - \hat{w}^{12} \hat{\chi}^{12} ) \gamma^{12} = 
\hat{w}^{12} \zeta^{12} + \vartheta^{12} ,
\label{sm_sumrule2}
\end{equation}
where $\hat{1} = 1 \otimes 1$ denotes the unit superoperator.

Let us introduce finally a symbol $\{ x \, ; y \}$, where $x$
and $y$ are arbitrary operators, which is defined by
\begin{equation}
\{ x \, ; y \} = {\rm Tr} (xy) .
\label{sm_trace}
\end{equation}
This symbol is symmetric, $\{ x \, ; y \} = \{ y \, ; x \}$,
linear in both arguments and it satisfies the rule
\begin{equation}
\{ ( a \otimes b ) x \, ; y \} = \{ x \, ; ( b \otimes a ) y \} ,
\label{sm_tr_rule1}
\end{equation}
which follows from the cyclic invariance of the trace. 
An obvious consequence of this rule are relations
\begin{eqnarray}
\{ \hat{w}^{12} x \, ; y \} & = & \{ x \, ; \hat{w}^{21} y \} ,
\nonumber\\
\{ \hat{\chi}^{12} x \, ; y \} & = & \{ x \, ; \hat{\chi}^{21} y \} ,
\label{sm_tr_rule2}
\end{eqnarray}
where $\hat{w}^{21}$ and $\hat{\chi}^{21}$ are defined by
(\ref{sm_w12_def}) and (\ref{sm_chi12_def}) with the superscript
interchange $1 \leftrightarrow 2$.

\subsection{Expression with the nonlocal torque}

The configuration averaging in $\beta^{\rm nl}$ (\ref{sm_beta_def}),
which contains the nonrandom operator $[K,S]$, leads to two terms 
\begin{equation}
\beta^{\rm nl} = \beta^{\rm nl,coh} + \beta^{\rm nl,vc} ,
\label{sm_beta_nl}
\end{equation}
where the coherent part is given by
\begin{equation}
\beta^{\rm nl,coh} = {\rm Tr} \{ \bgo [ K , S ] \bgt [ K , S ] \} 
\label{sm_beta_nl_coh_def}
\end{equation}
and the vertex corrections can be compactly written
as \cite{r_2006_ctk}
\begin{equation} 
\beta^{\rm nl,vc} = \{ ( \hat{1} - \hat{w}^{12} \hat{\chi}^{12} 
)^{-1} \hat{w}^{12} \zeta^{12} \, ; \zeta^{21} \} , 
\label{sm_beta_nl_vc_def}
\end{equation}
with all symbols and quantities defined in the previous section.
The coherent part can be written as a sum of four terms,
\begin{eqnarray}
\beta^{\rm nl,coh} & = & 
  \beta^{\rm nl,coh}_A + \beta^{\rm nl,coh}_B
+ \beta^{\rm nl,coh}_C + \beta^{\rm nl,coh}_D ,
\nonumber\\
\beta^{\rm nl,coh}_A & = & 
{\rm Tr} \{ S \bgo K S \bgt K \} ,
\nonumber\\
\beta^{\rm nl,coh}_B & = & 
{\rm Tr} \{ \bgo S K \bgt S K \} ,
\nonumber\\
\beta^{\rm nl,coh}_C & = & 
- {\rm Tr} \{ \bgo K S \bgt S K \} ,
\nonumber\\
\beta^{\rm nl,coh}_D & = & 
- {\rm Tr} \{ S \bgo S K \bgt K \} ,
\label{sm_beta_nl_coh_q}
\end{eqnarray}
which can be further modified using the equation of motion
(\ref{sm_eom}) and its consequences, e.g., $S {\bar g}^j =
{\cal P}^j {\bar g}^j - 1$.
For the first term $\beta^{\rm nl,coh}_A$, one obtains:
\begin{eqnarray}
\beta^{\rm nl,coh}_A & = & 
{\rm Tr} \{ {\cal P}^1 \bgo K {\cal P}^2 \bgt K \} 
+ {\rm Tr} \{ K K \} 
\nonumber\\ 
 & & {} - {\rm Tr} \{ K {\cal P}^2 \bgt K \}
        - {\rm Tr} \{ {\cal P}^1 \bgo K K \} .
\quad
\label{sm_beta_nl_coh_a}
\end{eqnarray}
The last three terms do not contribute to the sum over four
pairs of indices $(p,q)$, where $p, q \in \{ + , - \}$,
in Eq.~(\ref{sm_alpha_nl}).
For this reason, they can be omitted for the present purpose,
which yields expressions
\begin{eqnarray}
{\tilde \beta}^{\rm nl,coh}_A & = &
{\rm Tr} \{ {\cal P}^1 \bgo K {\cal P}^2 \bgt K \} ,
\nonumber\\
{\tilde \beta}^{\rm nl,coh}_B & = &
{\rm Tr} \{ \bgo {\cal P}^1 K \bgt {\cal P}^2 K \} ,
\label{sm_tilbeta_nl_coh_ab}
\end{eqnarray}
where the second relation is obtained in the same way from the
original term $\beta^{\rm nl,coh}_B$.
A similar approach can be applied to the third term
$\beta^{\rm nl,coh}_C$, which yields  
\begin{eqnarray}
\beta^{\rm nl,coh}_C & = & 
- {\rm Tr} \{ \bgo K {\cal P}^2 \bgt {\cal P}^2 K \} 
\nonumber\\ 
 & & {} + {\rm Tr} \{ \bgo K {\cal P}^2 K \}
        + {\rm Tr} \{ \bgo K S K \} .
\quad
\label{sm_beta_nl_coh_c}
\end{eqnarray}
The last term does not contribute to the sum over four pairs
$(p,q)$ in Eq.~(\ref{sm_alpha_nl}), which leads to expressions
\begin{eqnarray}
{\tilde \beta}^{\rm nl,coh}_C & = & 
  {\rm Tr} \{ \bgo K {\cal P}^2 K \}
- {\rm Tr} \{ \bgo K {\cal P}^2 \bgt {\cal P}^2 K \} ,
\nonumber\\
{\tilde \beta}^{\rm nl,coh}_D & = & 
  {\rm Tr} \{ {\cal P}^1 K \bgt K \}
- {\rm Tr} \{ {\cal P}^1 \bgo {\cal P}^1 K \bgt K \} ,
\qquad
\label{sm_tilbeta_nl_coh_cd}
\end{eqnarray}
where the second relation is obtained in the same way from the
original term $\beta^{\rm nl,coh}_D$.
The sum of all four contributions in (\ref{sm_tilbeta_nl_coh_ab})
and (\ref{sm_tilbeta_nl_coh_cd}) yields
\begin{eqnarray}
{\tilde \beta}^{\rm nl,coh} & = &
  {\tilde \beta}^{\rm nl,coh}_A + {\tilde \beta}^{\rm nl,coh}_B 
+ {\tilde \beta}^{\rm nl,coh}_C + {\tilde \beta}^{\rm nl,coh}_D 
\nonumber\\
 & = &
  {\rm Tr} \{ \bgo K {\cal P}^2 K \} 
+ {\rm Tr} \{ {\cal P}^1 K \bgt K \} 
\nonumber\\
 & & {} + {\rm Tr} \{ \bgo \gamma^{12} \bgt \gamma^{21} \} ,
\label{sm_tilbeta_nl_coh}
\end{eqnarray}
where we used the operators $\gamma^{12}$ and $\gamma^{21}$
defined by (\ref{sm_gamma_def}).
The total quantity $\beta^{\rm nl}$ (\ref{sm_beta_nl}) is thus
equivalent to
\begin{eqnarray}
{\tilde \beta}^{\rm nl} & = & 
{\tilde \beta}^{\rm nl,coh} + \beta^{\rm nl,vc} 
\nonumber\\
 & = &
  {\rm Tr} \{ \bgo K {\cal P}^2 K \}
+ {\rm Tr} \{ {\cal P}^1 K \bgt K \}
\nonumber\\
 & & {} + {\rm Tr} \{ \bgo \gamma^{12} \bgt \gamma^{21} \}
+ \beta^{\rm nl,vc} ,
\label{sm_tilbeta_nl}
\end{eqnarray}
where the tildes mark omission of terms irrelevant for the
summation over $(p,q)$ in Eq.~(\ref{sm_alpha_nl}). 

\subsection{Expression with the local torque}

The configuration averaging in $\beta^{\rm loc}$
(\ref{sm_beta_def}), involving the random local torque, leads to
a sum of two terms: \cite{r_1985_whb}
\begin{equation}
\beta^{\rm loc} = \beta^{\rm loc,0} + \beta^{\rm loc,1} ,
\label{sm_beta_loc}
\end{equation}
where the term $\beta^{\rm loc,0}$ is given by a simple lattice sum
\begin{eqnarray}
\beta^{\rm loc,0} & = & \sum_{\bf R} \beta^{\rm loc,0}_{\bf R} ,
\nonumber\\
\beta^{\rm loc,0}_{\bf R} & = & 
{\rm Tr} \left\langle \bgo ( 1 - \tmo \bgo ) ( \pfo K - K \pft )
\right. 
\nonumber\\
 & & \left. \ {} \times
\bgt ( 1 - \tmt \bgt ) ( \pft K - K \pfo ) \right\rangle , 
\quad
\label{sm_beta_loc0_def}
\end{eqnarray}
see Eq.~(76) of Ref.~\onlinecite{r_1985_whb}, and
the term $\beta^{\rm loc,1}$ can be written in the present
formalism as 
\begin{equation} 
\beta^{\rm loc,1} = \{ \hat{\chi}^{12} ( \hat{1} - \hat{w}^{12}
\hat{\chi}^{12} )^{-1} \vartheta^{12} \, ; \vartheta^{21} \} , 
\label{sm_beta_loc1_def}
\end{equation}
which corresponds to Eq.~(74) of Ref.~\onlinecite{r_1985_whb}.
The definitions of $\hat{w}^{12}$ and $\hat{\chi}^{12}$ are given
by (\ref{sm_w12_def}) and (\ref{sm_chi12_def}), respectively, and
of $\vartheta^{12}$ and $\vartheta^{21}$ by (\ref{sm_theta_def}). 

The quantity $\beta^{\rm loc,0}_{\bf R}$ (\ref{sm_beta_loc0_def})
gives rise to four terms,
\begin{eqnarray}
\label{sm_beta_loc0_r}
\beta^{\rm loc,0}_{\bf R} & = & 
Q_{{\bf R},A} + Q_{{\bf R},B} + Q_{{\bf R},C} + Q_{{\bf R},D} ,
 \\
Q_{{\bf R},A} & = & 
{\rm Tr} \langle \bgo ( 1 - \tmo \bgo ) \pfo K
\bgt ( 1 - \tmt \bgt ) \pft K \rangle ,
\nonumber\\
Q_{{\bf R},B} & = & 
{\rm Tr} \langle \pfo \bgo ( 1 - \tmo \bgo ) K
\pft \bgt ( 1 - \tmt \bgt ) K \rangle ,
\nonumber\\
Q_{{\bf R},C} & = & 
- {\rm Tr} \langle \pfo \bgo ( 1 - \tmo \bgo ) \pfo K
\bgt ( 1 - \tmt \bgt ) K \rangle ,
\nonumber\\
Q_{{\bf R},D} & = & 
- {\rm Tr} \langle \bgo ( 1 - \tmo \bgo ) K
\pft \bgt ( 1 - \tmt \bgt ) \pft K \rangle ,
\nonumber
\end{eqnarray}
which will be treated separately.
The term $Q_{{\bf R},A}$ can be simplified by employing the
identities (\ref{sm_tmid}) and the CPA-selfconsistency conditions.
This yields:
\begin{eqnarray}
\label{sm_beta_loc0_qa}
Q_{{\bf R},A} & = & U_{{\bf R},A} + V_{{\bf R},A} ,
\\
U_{{\bf R},A} & = &
{\rm Tr} \{ \bgo \cpo K \bgt \cpt K \} ,
\nonumber\\
V_{{\bf R},A} & = & {\rm Tr} \langle 
\bgo \tmo ( 1 - \bgo \cpo ) K \bgt \tmt ( 1 - \bgt \cpt ) K \rangle
\nonumber\\
 & = & 
V_{{\bf R},A1} + V_{{\bf R},A2} + V_{{\bf R},A3} +V_{{\bf R},A4} ,
\nonumber\\
V_{{\bf R},A1} & = & {\rm Tr} \langle \bgo \tmo K
\bgt \tmt K \rangle ,
\nonumber\\
V_{{\bf R},A2} & = & {\rm Tr} \langle \bgo \tmo \bgo \cpo K
\bgt \tmt \bgt \cpt K \rangle ,
\nonumber\\
V_{{\bf R},A3} & = & - {\rm Tr} \langle \bgo \tmo \bgo \cpo K
\bgt \tmt K \rangle ,
\nonumber\\
V_{{\bf R},A4} & = & - {\rm Tr} \langle \bgo \tmo K
\bgt \tmt \bgt \cpt K \rangle .
\nonumber
\end{eqnarray}
A similar procedure applied to $Q_{{\bf R},B}$ yields:
\begin{eqnarray}
\label{sm_beta_loc0_qb}
Q_{{\bf R},B} & = & U_{{\bf R},B} + V_{{\bf R},B} ,
\\
U_{{\bf R},B} & = &
{\rm Tr} \{ \cpo \bgo K \cpt \bgt K \} ,
\nonumber\\
V_{{\bf R},B} & = & {\rm Tr} \langle 
( 1 - \cpo \bgo ) \tmo \bgo K ( 1 - \cpt \bgt ) \tmt \bgt K \rangle
\nonumber\\
 & = & 
V_{{\bf R},B1} + V_{{\bf R},B2} + V_{{\bf R},B3} +V_{{\bf R},B4} ,
\nonumber\\
V_{{\bf R},B1} & = & {\rm Tr} \langle \tmo \bgo K
\tmt \bgt K \rangle ,
\nonumber\\
V_{{\bf R},B2} & = & {\rm Tr} \langle \cpo \bgo \tmo \bgo K
\cpt \bgt \tmt \bgt K \rangle ,
\nonumber\\
V_{{\bf R},B3} & = & - {\rm Tr} \langle \tmo \bgo K
\cpt \bgt \tmt \bgt K \rangle ,
\nonumber\\
V_{{\bf R},B4} & = & - {\rm Tr} \langle \cpo \bgo \tmo \bgo K
\tmt \bgt K \rangle .
\nonumber
\end{eqnarray}
The term $Q_{{\bf R},C}$ requires an auxiliary relation
\begin{eqnarray}
\pfo \bgo ( 1 - \tmo \bgo ) \pfo = 
\cpo ( \bgo \cpo - 1 ) \phantom{m} & &
\nonumber\\
{} + \pfo - ( 1 - \cpo \bgo ) \tmo ( 1 - \bgo \cpo ) , & & 
\label{sm_beta_loc0_paux}
\end{eqnarray}
that follows from a repeated use of the identities (\ref{sm_tmid}).
This relation together with the CPA-selfconsistency lead to the
form:
\begin{eqnarray}
\label{sm_beta_loc0_qc}
Q_{{\bf R},C} & = & U_{{\bf R},C} + V_{{\bf R},C} ,
\\
U_{{\bf R},C} & = &
{\rm Tr} \{ \cpo ( 1 - \bgo \cpo ) K \bgt K \}
\nonumber\\
 & & 
 {} - {\rm Tr} \langle \pfo K \bgt ( 1 - \tmt \bgt ) K \rangle ,
\nonumber\\
V_{{\bf R},C} & = & - {\rm Tr} \langle 
( 1 - \cpo \bgo ) \tmo ( 1 - \bgo \cpo ) K \bgt \tmt \bgt K \rangle
\nonumber\\
 & = & 
V_{{\bf R},C1} + V_{{\bf R},C2} + V_{{\bf R},C3} +V_{{\bf R},C4} ,
\nonumber\\
V_{{\bf R},C1} & = & - {\rm Tr} \langle \tmo K
\bgt \tmt \bgt K \rangle ,
\nonumber\\
V_{{\bf R},C2} & = & - {\rm Tr} \langle \cpo \bgo \tmo \bgo \cpo K
\bgt \tmt \bgt K \rangle ,
\nonumber\\
V_{{\bf R},C3} & = & {\rm Tr} \langle \tmo \bgo \cpo K
\bgt \tmt \bgt K \rangle ,
\nonumber\\
V_{{\bf R},C4} & = & {\rm Tr} \langle \cpo \bgo \tmo K
\bgt \tmt \bgt K \rangle .
\nonumber
\end{eqnarray}
A similar procedure applied to $Q_{{\bf R},D}$ yields:
\begin{eqnarray}
\label{sm_beta_loc0_qd}
Q_{{\bf R},D} & = & U_{{\bf R},D} + V_{{\bf R},D} ,
\\
U_{{\bf R},D} & = &
{\rm Tr} \{ \bgo K \cpt ( 1 - \bgt \cpt ) K \}
\nonumber\\
 & & 
 {} - {\rm Tr} \langle \bgo ( 1 - \tmo \bgo ) K \pft K \rangle ,
\nonumber\\
V_{{\bf R},D} & = & - {\rm Tr} \langle 
\bgo \tmo \bgo K ( 1 - \cpt \bgt ) \tmt ( 1 - \bgt \cpt ) K \rangle
\nonumber\\
 & = & 
V_{{\bf R},D1} + V_{{\bf R},D2} + V_{{\bf R},D3} +V_{{\bf R},D4} ,
\nonumber\\
V_{{\bf R},D1} & = & - {\rm Tr} \langle \bgo \tmo \bgo K
\tmt K \rangle ,
\nonumber\\
V_{{\bf R},D2} & = & - {\rm Tr} \langle \bgo \tmo \bgo K
\cpt \bgt \tmt \bgt \cpt K \rangle ,
\nonumber\\
V_{{\bf R},D3} & = & {\rm Tr} \langle \bgo \tmo \bgo K
\cpt \bgt \tmt K \rangle ,
\nonumber\\
V_{{\bf R},D4} & = & {\rm Tr} \langle \bgo \tmo \bgo K
\tmt \bgt \cpt K \rangle ,
\nonumber
\end{eqnarray}
Let us focus now on $U$-terms in Eqs.~(\ref{sm_beta_loc0_qa}
-- \ref{sm_beta_loc0_qd}).
The second terms in $U_{{\bf R},C}$ (\ref{sm_beta_loc0_qc}) and
$U_{{\bf R},D}$ (\ref{sm_beta_loc0_qd}) do not contribute to the
sum over four pairs $(p,q)$ in Eq.~(\ref{sm_alpha_loc}), so that
the original $U_{{\bf R},C}$ and $U_{{\bf R},D}$ can be replaced
by equivalent expressions
\begin{eqnarray}
{\tilde U}_{{\bf R},C} & = &
{\rm Tr} \{ \cpo ( 1 - \bgo \cpo ) K \bgt K \} ,
\nonumber\\
{\tilde U}_{{\bf R},D} & = &
{\rm Tr} \{ \bgo K \cpt ( 1 - \bgt \cpt ) K \} .
\label{sm_beta_loc0_tucd}
\end{eqnarray}
The sum of all $U$-terms for the site ${\bf R}$ is then equal to
\begin{eqnarray}
{\tilde U}_{\bf R} & = & U_{{\bf R},A} + U_{{\bf R},B} 
+ {\tilde U}_{{\bf R},C} + {\tilde U}_{{\bf R},D} 
\nonumber\\
 & = & {\rm Tr} \{ \cpo K \bgt K \} + {\rm Tr} \{ \bgo K \cpt K \}
\nonumber\\
 & & {} + {\rm Tr} \{ \bgo \gamma^{12}_{\bf R} \bgt 
\gamma^{21}_{\bf R} \} ,
\label{sm_beta_loc0_ur}
\end{eqnarray}
where $\gamma^{12}_{\bf R}$ and $\gamma^{21}_{\bf R}$ are defined
in (\ref{sm_gamma_def}), and the lattice sum of all $U$-terms can be
written as
\begin{eqnarray}
\sum_{\bf R} {\tilde U}_{\bf R} & = & {\rm Tr} \{ 
{\cal P}^1 K \bgt K \} + {\rm Tr} \{ \bgo K {\cal P}^2 K \}
\nonumber\\
 & & {} + \sum_{\bf R} {\rm Tr} \{ \bgo \gamma^{12}_{\bf R}
\bgt \gamma^{21}_{\bf R} \} .
\label{sm_beta_loc0_usum}
\end{eqnarray}
The summation of $V$-terms in Eqs.~(\ref{sm_beta_loc0_qa}
-- \ref{sm_beta_loc0_qd}) can be done in two steps.
First, we obtain
\begin{eqnarray}
V_{{\bf R},1} & = & V_{{\bf R},A1} 
 + V_{{\bf R},B1} + V_{{\bf R},C1} + V_{{\bf R},D1} 
\nonumber\\
 & = & {\rm Tr} \langle \tmo f^{12} \tmt f^{21} \rangle ,
\nonumber\\
V_{{\bf R},2} & = & V_{{\bf R},A2} 
 + V_{{\bf R},B2} + V_{{\bf R},C2} + V_{{\bf R},D2} 
\nonumber\\
 & = & {\rm Tr} \langle \tmo \bgo \gamma^{12}_{\bf R}
\bgt \tmt \bgt \gamma^{21}_{\bf R} \bgo \rangle ,
\nonumber\\
V_{{\bf R},3} & = & V_{{\bf R},A3} 
 + V_{{\bf R},B3} + V_{{\bf R},C3} + V_{{\bf R},D3} 
\nonumber\\
 & = & {\rm Tr} \langle \tmo \bgo \gamma^{12}_{\bf R}
\bgt \tmt f^{21} \rangle ,
\nonumber\\
V_{{\bf R},4} & = & V_{{\bf R},A4} 
 + V_{{\bf R},B4} + V_{{\bf R},C4} + V_{{\bf R},D4} 
\nonumber\\
 & = & {\rm Tr} \langle \tmo f^{12} \tmt \bgt 
\gamma^{21}_{\bf R} \bgo \rangle ,
\label{sm_beta_loc0_vr1234}
\end{eqnarray}
where the operators $f^{12}$ and $f^{21}$ have been defined
in (\ref{sm_f_zeta_def}).
Second, one obtains the sum of all $V$-terms for the site 
${\bf R}$ as 
\begin{eqnarray}
\label{sm_beta_loc0_vr}
V_{\bf R} & = & V_{{\bf R},1} 
 + V_{{\bf R},2} + V_{{\bf R},3} + V_{{\bf R},4} 
\\
 & = & {\rm Tr} \langle 
\tmo ( f^{12} + \bgo \gamma^{12}_{\bf R} \bgt )
\tmt ( f^{21} + \bgt \gamma^{21}_{\bf R} \bgo )
\rangle .
\nonumber
\end{eqnarray}
The lattice sums of all $U$- and $V$-terms lead to an expression
equivalent to the original quantity $\beta^{\rm loc,0}$
(\ref{sm_beta_loc0_def}):  
\begin{eqnarray}
{\tilde \beta}^{\rm loc,0} & = & 
\sum_{\bf R} {\tilde U}_{\bf R} + \sum_{\bf R} V_{\bf R} 
\nonumber\\
 & = & {\rm Tr} \{ 
{\cal P}^1 K \bgt K \} + {\rm Tr} \{ \bgo K {\cal P}^2 K \}
\nonumber\\
 & & {} + \sum_{\bf R} {\rm Tr} \{ \bgo \gamma^{12}_{\bf R}
\bgt \gamma^{21}_{\bf R} \} 
\nonumber\\
 & & {} + \sum_{\bf R} {\rm Tr} \left\langle 
\tmo ( f^{12} + \bgo \gamma^{12}_{\bf R} \bgt ) \right.
\nonumber\\
 & & \left. \qquad \quad \, {} \times 
\tmt ( f^{21} + \bgt \gamma^{21}_{\bf R} \bgo )
\right\rangle ,
\label{sm_tilbeta_loc0}
\end{eqnarray}
where the tildes mark omission of terms not contributing to the
summation over $(p,q)$ in Eq.~(\ref{sm_alpha_loc}). 

Let us turn now to the contribution $\beta^{\rm loc,1}$
(\ref{sm_beta_loc1_def}).
It can be reformulated by expressing the quantity
$\vartheta^{12}$ (and $\vartheta^{21}$) in terms of the
quantities $\gamma^{12}$ and $\zeta^{12}$ 
(and $\gamma^{21}$ and $\zeta^{21}$) from the sum rule
(\ref{sm_sumrule2}) and by using the identities (\ref{sm_tr_rule2}).
The resulting form can be written compactly with help of an
auxiliary operator $\varrho^{12}$ (and $\varrho^{21}$) defined as
\begin{equation}
\varrho^{12} = \hat{\chi}^{12} \gamma^{12} + \zeta^{12} .
\label{sm_rho12_def}
\end{equation}
The result is
\begin{eqnarray}
\beta^{\rm loc,1} & = & \beta^{\rm nl,vc} 
 + \{ \hat{\chi}^{12} \gamma^{12}  \, ; \gamma^{21} \}
\nonumber\\
 & & {} - \{ \hat{w}^{12} \varrho^{12}  \, ; \varrho^{21} \} ,
\label{sm_beta_loc1_p}
\end{eqnarray}
where the first term has been defined in (\ref{sm_beta_nl_vc_def}).
For the second term in (\ref{sm_beta_loc1_p}), we use the relation
\begin{equation}
\hat{\chi}^{12} \gamma^{12} = \sum_{\bf R} \Pi_{\bf R} \bgo
( \gamma^{12} - \gamma^{12}_{\bf R} ) \bgt \Pi_{\bf R} ,
\label{sm_chigamma}
\end{equation}
which follows from the site-diagonal nature of the operator
$\gamma^{12}$ (\ref{sm_gamma_def}) and from the definition
of the superoperator $\hat{\chi}^{12}$ (\ref{sm_chi12_def}).
This yields:
\begin{eqnarray}
\{ \hat{\chi}^{12} \gamma^{12}  \, ; \gamma^{21} \} & = &
{\rm Tr} \{ \bgo \gamma^{12} \bgt \gamma^{21} \} 
\nonumber\\
 & & {} - \sum_{\bf R} {\rm Tr} 
\{ \bgo \gamma^{12}_{\bf R} \bgt \gamma^{21}_{\bf R} \} .
\label{sm_beta_loc1_term2}
\end{eqnarray}
For the third term in (\ref{sm_beta_loc1_p}), only the site-diagonal
blocks of the operator $\varrho^{12}$ (and $\varrho^{21}$), 
Eq.~(\ref{sm_rho12_def}), are needed because of the site-diagonal
nature of the superoperator $\hat{w}^{12}$ (\ref{sm_w12_def}). 
These site-diagonal blocks are given by
\begin{eqnarray}
\Pi_{\bf R} \varrho^{12} \Pi_{\bf R} & = & 
\Pi_{\bf R} \left[ \bgo ( \gamma^{12} - \gamma^{12}_{\bf R} )
 \bgt + \zeta^{12} \right] \Pi_{\bf R} 
\nonumber\\
 & = & 
- \Pi_{\bf R} ( f^{12} + \bgo \gamma^{12}_{\bf R} \bgt ) 
\Pi_{\bf R} ,
\label{sm_rho12_sdb}
\end{eqnarray}
which follows from the previous relations (\ref{sm_chigamma}) and
(\ref{sm_f_zeta_gamma}).

\noindent
This yields:
\begin{eqnarray}
 \{ \hat{w}^{12} \varrho^{12}  \, ; \varrho^{21} \} & = &
 \sum_{\bf R} {\rm Tr} \left\langle 
\tmo ( f^{12} + \bgo \gamma^{12}_{\bf R} \bgt ) \right.
\nonumber\\
 & & \left. \qquad {} \times 
\tmt ( f^{21} + \bgt \gamma^{21}_{\bf R} \bgo ) \right\rangle . 
\qquad
\label{sm_beta_loc1_term3}
\end{eqnarray}
The term $\beta^{\rm loc,1}$ (\ref{sm_beta_loc1_p}) is then
equal to 
\begin{eqnarray}
\beta^{\rm loc,1} & = & \beta^{\rm nl,vc} + 
{\rm Tr} \{ \bgo \gamma^{12} \bgt \gamma^{21} \} 
\nonumber\\
 & & {} - \sum_{\bf R} {\rm Tr} 
\{ \bgo \gamma^{12}_{\bf R} \bgt \gamma^{21}_{\bf R} \} 
\nonumber\\
 & & {} - \sum_{\bf R} {\rm Tr} \left\langle 
\tmo ( f^{12} + \bgo \gamma^{12}_{\bf R} \bgt ) \right.
\nonumber\\
 & & \left. \qquad \quad \, {} \times 
\tmt ( f^{21} + \bgt \gamma^{21}_{\bf R} \bgo ) \right\rangle . 
\label{sm_beta_loc1_fin}
\end{eqnarray}
The total quantity $\beta^{\rm loc}$ (\ref{sm_beta_loc}) is thus
equivalent to the sum of (\ref{sm_tilbeta_loc0}) and 
(\ref{sm_beta_loc1_fin}):
\begin{eqnarray}
{\tilde \beta}^{\rm loc} & = &
{\tilde \beta}^{\rm loc,0} + \beta^{\rm loc,1} 
\nonumber\\
 & = & {\rm Tr} \{ 
{\cal P}^1 K \bgt K \} + {\rm Tr} \{ \bgo K {\cal P}^2 K \}
\nonumber\\
 & & {} + {\rm Tr} \{ \bgo \gamma^{12} \bgt \gamma^{21} \} 
+ \beta^{\rm nl,vc} , 
\label{sm_tilbeta_loc}
\end{eqnarray}
where the tildes mark omission of terms irrelevant for the
summation over $(p,q)$ in Eq.~(\ref{sm_alpha_loc}). 

\vspace{5ex}

\subsection{Comparison of both expressions}

A comparison of relations (\ref{sm_tilbeta_nl}) and
(\ref{sm_tilbeta_loc}) shows immediately that
\begin{equation}
{\tilde \beta}^{\rm nl} = {\tilde \beta}^{\rm loc} , 
\label{sm_tilbeta_comp}
\end{equation}
which means that the original expressions $\beta^{\rm loc}$
and $\beta^{\rm nl}$ in (\ref{sm_beta_def}) are identical
up to terms not contributing to the $(p,q)$-summations
in (\ref{sm_alpha_loc}) and (\ref{sm_alpha_nl}).
This proves the exact equivalence of the Gilbert damping parameters
obtained with the local and nonlocal torques in the CPA.


\providecommand{\noopsort}[1]{}\providecommand{\singleletter}[1]{#1}%

\end{document}